\documentclass[prb,twocolumn,showpacs,preprintnumbers,amsmath,aps]{revtex4}
\usepackage{graphicx,hyperref}
\usepackage{color}
\usepackage{bm}
\usepackage{subfigure}

\begin{document}

\title{Interplay of vitrification and ice formation in a cryoprotectant aqueous solution at low temperature}

\author{Christiane Alba-Simionesco}
\email{christiane.alba-simionesco@cea.fr}
\author{Patrick Judeinstein, St\'ephane Longeville, Oriana Osta, Florence Porcher}
\affiliation{University Paris-Saclay, CEA, CNRS, Laboratoire L\'eon Brillouin, 91191, Gif-sur-Yvette, France}

\author{Fr\'ed\'eric Caupin} 
\affiliation{Institut Lumi\`ere Mati\`ere, CNRS, University Claude Bernard-Lyon 1, 6 rue Ada Byron, France}

\author{Gilles Tarjus} \email{tarjus@lptmc.jussieu.fr}
\affiliation{LPTMC, CNRS, Sorbonne Universit\'e, 4 Place Jussieu, 75005 Paris, France}

\date{\today}

\begin{abstract}
The proneness of water to crystallize is a major obstacle to understanding its putative exotic behavior in the supercooled state. It also 
represents a strong practical limitation to cryopreservation of biological systems. Adding some concentration of glycerol, which has a 
cryoprotective effect preventing to some degree water crystallization, has been proposed as a possible way out, provided the concentration is small 
enough for water to retain some of its bulk character and/or for limiting the damage caused by glycerol on living organisms. Contrary to previous expectations, 
we show that in the ``marginal'' glycerol molar concentration $\approx18\%$, at which vitrification is possible with no crystallization on rapid cooling, 
water crystallizes upon isothermal annealing even below the calorimetric glass transition of the solution. Through a time-resolved polarized neutron scattering 
investigation, we extract key parameters, size and shape of the ice crystallites, fraction of water that crystallizes, crystallization time, 
which are important for cryoprotection, as a function of the annealing temperature. We also characterize the nature of the out-of-equilibrium liquid phases 
that are present at low temperature, providing more arguments against the presence of an iso-compositional liquid-liquid transition. Finally, we propose a 
rule-of-thumb to estimate the lower temperature limit below which water crystallization does not occur in aqueous solutions.
\end{abstract}
\maketitle

{\bf Significance statement: 
Studying water crystallization at low temperature and the lower limit of ice formation is crucial both for a fundamental understanding of water and for practical 
reasons such as cryopreservation. By taking advantage of the polarized neutron scattering technique and by considering a nano-segregated water-glycerol solution 
we are able to characterize the key parameters of ice formation at temperatures near and below the calorimetric glass transition of the solution and provide a general 
rule for estimating the lower temperature limit of water crystallization in a broad range of aqueous solutions. We also show that nano-segregated 
water in the glassy solution at low temperature is not in a high-density form but in a low-density one.}
\\

Adding glycerol to water is known to inhibit ice formation because of the perturbation that glycerol produces on the hydrogen-bonding network of water. This 
property has important consequences both on a practical and on a theoretical level. Glycerol is one example of a cryoprotectant, cryoprotectants being chemicals 
used to protect biological molecules, organs, plants and insects from 
freezing\cite{poldge49,farrant65,fahy84,fahy09,zachariassen_review,cryop_review,cryop_plants_review,stachecki19,fahy21}. Its addition to an aqueous solution 
may even allow its vitrification, provided a fast enough cooling protocol is applied, thereby opening the possibility of long-term preservation at low temperature of 
cells, plants, or at a different level protein structures\cite{protein}, without the damaging interference of ice formation. On the other hand, being able 
to get around crystallization of water that is otherwise unavoidable in a temperature range between $235$ K and around $150$ K is a route to 
study the exciting and actively debated properties of the putative supercooled liquid water in this range\cite{mishima_H2O,poole92,debenedetti03}. 

In both cases the concentration $c_g$ of glycerol must somehow be optimized. Indeed, a too small concentration is not sufficient to prevent water crystallization by 
rapid but standard cooling techniques, whereas a too high concentration strongly perturbs water in the mixture which may then lose its resemblance with bulk water; 
such a high concentration then invalidates the theoretical value of the mixture as a proxy for bulk water and, on the other hand, may damage the living organisms 
that one is trying to preserve\cite{poldge49,fahy84,zachariassen_review,cryop_review,cryop_plants_review,stachecki19,fahy21}. So, one is especially interested 
in the lowest glycerol concentration for which glass formation is still possible by rapid cooling, say in liquid nitrogen. It has been shown to be at least $15\%$  in molar 
concentration\cite{inaba07,suzuki14,popov15,loerting16}. We have then chosen a concentration of about $18\%$ for which, in addition, previously obtained data was 
available. (Here and in what follows we use the molar concentration $c_g$; $c_g=0.18$ then corresponds to a mass concentration of about $52$-$52.5\%$ 
depending on the deuteration.)

Our goal is two-fold. First, we want to assess for the ``marginal'' concentration $c_g\approx 0.18$ at which vitrification is possible with no crystallization upon rapid 
cooling the factors that characterize water crystallization and determine the cryoprotective ability of glycerol. The glass state is out of equilibrium and, depending 
on how deep in it the solution is, it may undergo some form of relaxation and age, which could in rare instances lead to very slow crystallization. Relevant questions 
that have not been addressed so far are thus: What is the timescale for crystallization in this cryoprotectant aqueous solution at low temperature and how does it vary with temperature? What are the characteristics of ice formation when the solution is in the vicinity of its calorimetric glass transition temperature? Is there a lowest 
temperature below which ice can no longer appear?
Second, we revisit the proposal made by Tanaka and coworkers\cite{murata12,murata13} that an ``iso-compositional'' liquid-liquid transition of the solution, 
triggered by an underlying liquid-liquid transition of water between a low-density and a high-density phase, takes place for $c_g\approx 0.18$. Although already 
criticized and, in our opinion, convincingly refuted by several authors\cite{suzuki14,popov15,loerting16}, the proposal has been very recently asserted 
again\cite{tanaka20} and it seems timely to add relevant experimental facts to the debate. We thus characterize the liquid phases appearing below melting, 
emphasizing their out-of-equilibrium character. We show in particular that nano-segregated water in the glassy solution 
at low temperature is not in a high-density  amorphous form but rather in a low-density one. 

Our study enables us to describe the stages and the main properties of slow ice formation in a glassy environment. This may be of interest for a 
better understanding of ice under astrophysical conditions (comets, planets, and interstellar matter)\cite{jenniskens96}. It also pertains 
to the broader scope of water polyamorphism and crystallization in electrolytic or nonelectrolytic aqueous solutions, a topic that has been 
extensively studied (for reviews see, e.g., [\onlinecite{angell_aqueous,loerting19}]). In particular, based on the temperature dependence of 
the  typical crystallite size that we obtain, we propose a practical way to estimate the limiting temperature below which water crystallization cannot occur 
for a range of aqueous solutions. The only required piece of information on the solution is its calorimetric glass-transition temperature. 

Before any further exposition, it is worth recalling the overall temperature-concentration phase diagram of water-glycerol 
solutions\cite{lane25,harran78,kanno05,inaba07,murata12,popov15,loerting16,niss18,nakagawa19,morineau20}: see Fig. S1 in the Supporting Information (SI). It is 
reasonably well established that below melting, three different ranges of glycerol concentration should be distinguished. At low concentration, $0<c_g\lesssim 0.15$, 
the presence of glycerol is not sufficient to prevent crystallization of water even by a deep quench in liquid nitrogen (alternative techniques should then be 
used: see, {\it e.g.}, [\onlinecite{suzuki20}]). At high concentration, $0.28\lesssim c_g<1$, the cryoprotective effect of glycerol controls the thermodynamic behavior  
and ice formation is easily avoided by a fast cooling; water molecules are well miscible with glycerol and no significant phase separation takes place. Above the 
so-called maximally-freeze concentrated solution $c_g\approx 0.38$, it is even virtually impossible to crystallize the solution, no matter how slow the cooling rate, 
except by introducing a crystal seed\cite{lane25}. Finally, the intermediate range, $0.15<c_g \lesssim 0.28$, is the more complex and interesting one, in which 
a strong dependence on the thermal treatments is found. Water crystallization can be avoided by fast cooling but is then observed upon heating. It is also a range where 
nano-segregation may play an important role. 

The core of our study is a time-resolved structural characterization of the phases and phase transformations of a water-glycerol mixture obtained after a fast 
quench at low temperature. This is done mostly at the glycerol concentration $c_g= 0.178\pm0.005$ through polarized neutron scattering  
and selective deuterations. An extensive set of data already exists for this concentration or nearby 
ones\cite{harran78,hayashi05,inaba07,murata12,suzuki14,popov15,loerting16,zhao15,bruijn16}, but no detailed structural investigations have been provided so far. 
Through the time-resolved neutron-scattering experiment we are able to probe the kinetics of phase transformation at constant annealing temperature, which may 
be very long in the glassy regime ($10$ hours or more). We further complement our analysis by thermodynamic measurements made by differential scanning 
calorimetry (DSC) and dynamical measurements obtained by Nuclear Magnetic Resonance (NMR) and Neutron Spin Echo (NSE).

\section{Results}
\label{Results}

\begin{figure}[tbp]
\centerline{\includegraphics[width=1\linewidth]{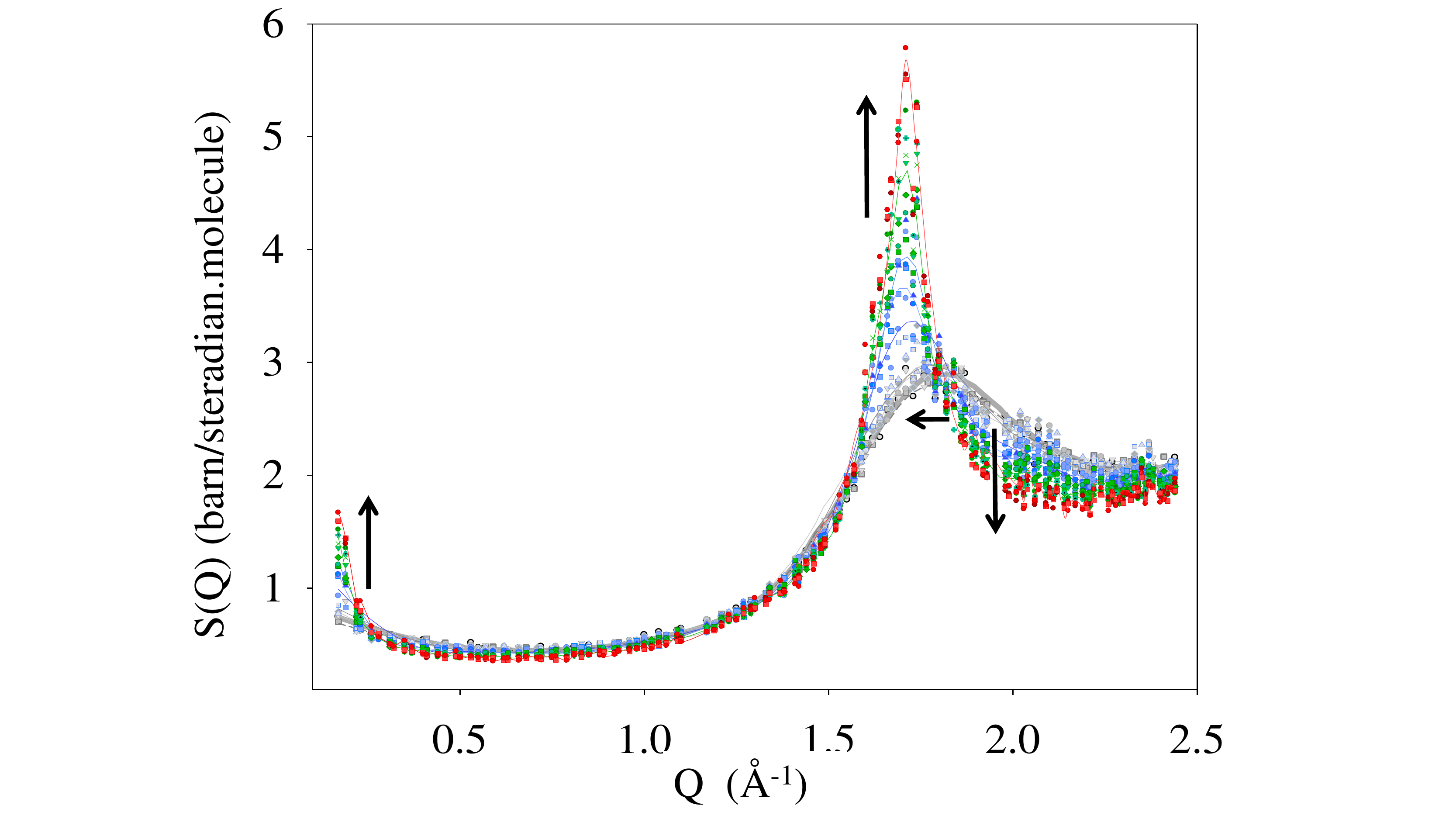}}
\centerline{\includegraphics[width=1\linewidth]{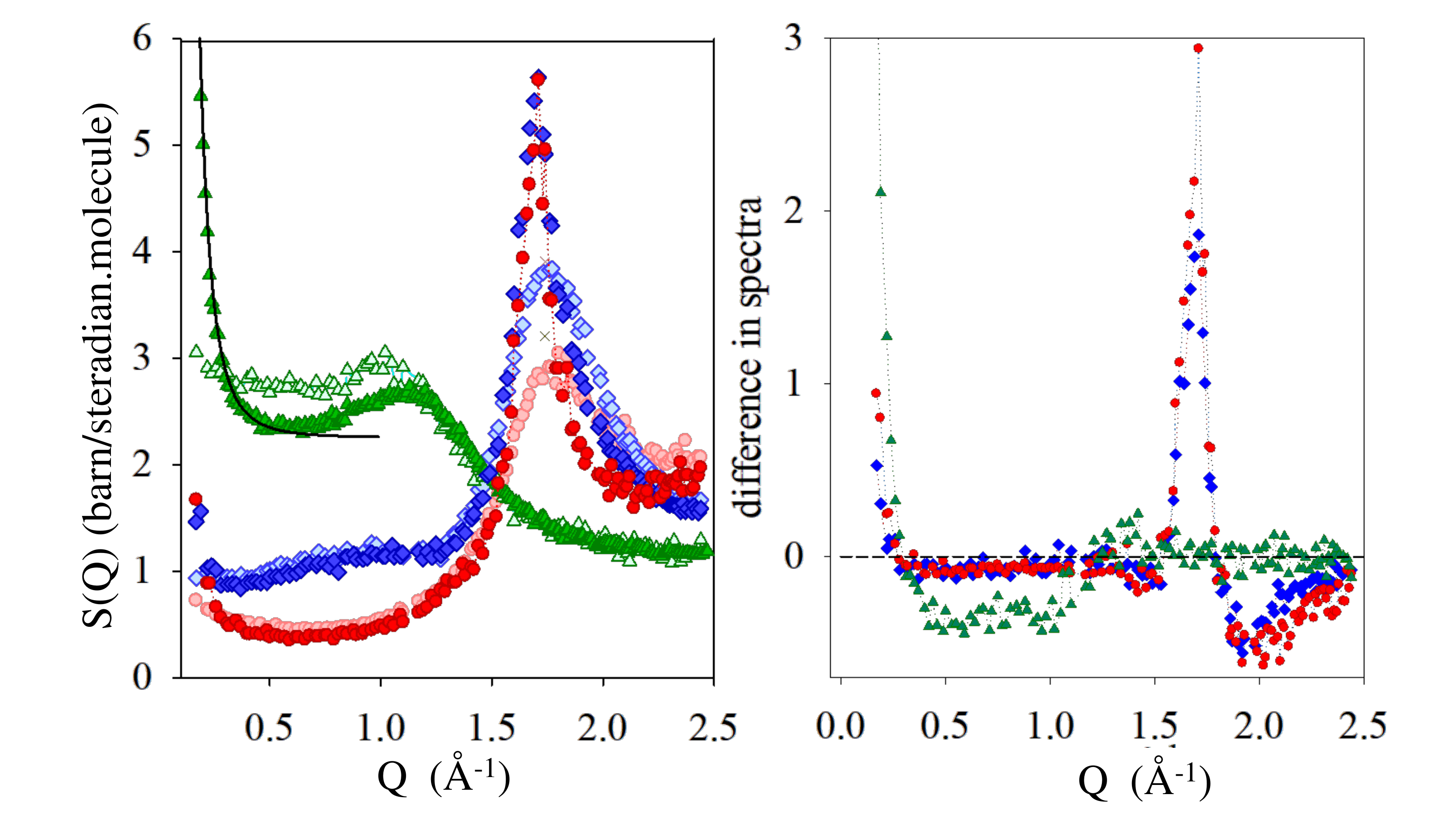}}
\caption{Top: Evolution of the static structure factor $S(Q)$ of the fully deuterated $17.8\%$ glycerol rich aqueous solution during an isothermal annealing at 
$T=160.2$ K, slightly below the calorimetric glass transition $T_g\approx 164.7$ K. Data are colored according to 4 domains of time: gray for $0$ to $176$ min, 
then blue up to $366$ min, green up to $620$ min and finally red up to $721$ min. The arrows indicate the main changes with time. 
Bottom left: Static structure factor $S(Q)$ of the fully deuterated ${\rm C_3D_5(OD)_3+D_2O}$ (red symbols) and partially deuterated ${\rm 
C_3D_5(OH)_3+H_2O}$ (green symbols) and ${\rm C_3H_5(OD)_3+D_2O}$ (blue symbols) $c_g= 0.178$ solution at the beginning (light color) and 
the end (dark color) of an isothermal annealing at $T=160.2$ K; the dark line shows the expected $Q^{-4}$ Porod's law at low $Q$ (an 
analysis is provided in Sec.~VI of the SI). Bottom right: same data shown as a difference between the final and the initial curves.}
\label{Fig_evolutionS(Q)_160}
\end{figure}

{\bf Water crystallization near and below the calorimetric glass transition of the ``marginal'' cryoprotectant solution}

{\it Evidence.} The glass transition temperature of the fully deuterated aqueous solution with $18\%$ of glycerol is found by DSC (cooling and heating rates of 
$10$ K/min) at $T_g\approx 164.7$ K, {\it i.e.}, is about $8$-$10$ K above that of the corresponding fully hydrogenated solution: see Fig.~S1 of the SI. 
(Generically, the characteristic temperatures of the deuterated sample are several degrees higher than those of the hydrogenated sample, e.g., about $4$ K higher 
for  the melting temperature $T_m$\cite{footnote_quantum}.) We study the evolution of the static structure factor $S(Q)$ of the solution during the annealing at 
$160$ K, {\it i.e.}, slightly below $T_g$, and at $170$ K, slightly above $T_g$. Note that due to the fast quench down to $80$-$90$ K ($\sim 70$-$130$ K/min) 
there is no sign of crystallization in the glass phase up to $160$ K prior to annealing. (This is discussed in Sec. III of the SI. In addition, in the SI we 
also provide an account of the effect of changing the cooling protocol on the results.)

We find that water crystallization takes place at $160$ K, albeit at a very slow pace. Evidence is shown in Fig.~\ref{Fig_evolutionS(Q)_160} where the isothermal 
evolution over time of the static structure factor $S(Q)$ of the fully deuterated sample is plotted. The distinctive signatures of ice formation are as follows: (i) a shift 
of the location  of the main peak to a lower wavevector, from $\approx1.75\AA^{-1}$ to $1.71\AA^{-1}$, (ii) a steep increase of the peak with a  concomitant narrowing, 
(iii) an increase at the lowest $Q$ values, and (iv) a strong decrease of the scattering intensity above the main-peak position, around $2\AA^{-1}$. This is even more 
clearly seen by looking at the difference between the data at the beginning and at the end of the annealing that is displayed in the bottom right panel. These features can 
all be interpreted as resulting from partial crystallization of water: (i)-(ii) the peak at $1.71\AA^{-1}$ matches  
the main peak of cubic ice $I_c$ and corresponds to a broadened Bragg peak that grows as more water crystallizes; (iii) the rise at low $Q$ is typical of the 
formation of interfaces, here between ice crystallites and the remaining liquid phase, and is described by Porod's $Q^{-4}$ law; finally, (iv) the depletion around 
$2\AA^{-1}$ is due to a decrease in the spatial correlations between water molecules and the alkyl chains of glycerol resulting from the formation of water 
crystallites. This decrease already appears when quenching the liquid into the glass due to a nano-segregation leading to the formation of domains of pure 
water, but it is much more developed when water crystallizes in the domains and more water aggregates to form ice. 

To complement this characterization we also display the structure factor $S(Q)$ at the beginning and the end of the annealing for 
two partially  deuterated samples, ${\rm C_3D_5(OH)_3+H_2O}$, ${\rm C_3H_5(OD)_3+D_2O}$, together with the fully deuterated one, 
${\rm C_3D_5(OD)_3+D_2O}$. The $S(Q)$ of ${\rm C_3D_5(OH)_3+H_2O}$ allows one to focus on the alkyl chains of glycerol, which, as seen from the Figure, 
tend to become closer as water crystallizes, with a peak in $S(Q)$ moving from about $1.05\AA^{-1}$ to $1.2\AA^{-1}$. One can also very clearly see the 
upswing at small $Q$'s due to the $Q^{-4}$ Porod law associated with the interfaces formed by ice crystallites (see the analysis in Sec.~VI of the SI). 
The structure factor of ${\rm C_3H_5(OD)_3+D_2O}$, for which all hydrogen atoms are deuterated except those of the alkyl chains, shows similar features to 
that of the fully deuterated sample. In particular, the three signatures of water crystallization, namely, the shift, growth and narrowing of the main peak, 
the upswing at the lowest $Q$'s and the intensity decrease in the region around $2\AA^{-1}$, can be clearly seen. There are some differences between 
the curves of this partially deuterated sample and the fully deuterated one: This stems from differences in the prefactors weighting the contributions 
of the partial structure factors to the total $S(Q)$ obtained by neutron scattering; in particular, the prefactors 
involving hydrogen atoms are mostly negative, which,  because of the overall normalization, mechanically leads to an increase of the prefactors weighting 
the other contributions.

All of this shows that water crystallization takes place at low temperature just below the glass transition in the presence of a molar concentration of $18\%$ of 
cryoprotectant glycerol, even after a rapid quench during which no crystallization occurred. This is the main finding of our study, on which we further elaborate 
below. This crystallization is a consequence of the nano-segregation of the solution that leads to the formation of small domains of water, a phenomenon whose premises  
can already be found in the liquid structure near the melting temperature\cite{soper16}. Nano-segregation of water, and its connection to ice formation,  
is a more general phenomenon that has also been observed for instance in aqueous salt solutions for a range of concentrations\cite{le-molinero11,bullock-molinero13,lane20}.

We have repeated the analysis for an annealing temperature of $170$ K, slightly above the glass transition temperature. We also observe water crystallization,  
and the evolution is now significantly faster than at $160$ K (see below). Water crystallization again appears as a first-order-like transition, as further 
supported by the presence of an exothermic peak in the DSC measurement
: see Sec.~VII of the SI. Evidence for water (cold) 
crystallization at this temperature or at a higher one has already been obtained by previous authors upon heating, by 
DSC\cite{harran78,hayashi05,popov15,loerting16}, adiabatic calorimetry\cite{inaba07}, X-ray scattering\cite{murata12}, dielectric spectroscopy\cite{popov15}, 
Raman scattering\cite{murata12,suzuki14} and infrared spectroscopy\cite{bruijn16}.

We note that the $S(Q)$'s obtained by either first annealing at $160$ K followed by heating at $170$ K or annealing directly at $170$ K are identical 
(see Fig. S7 of the SI), a robustness which confirms that metastable states are reached at the end of the annealing times, with no further water 
crystallization. After heating the sample(s) to still higher temperature and taking measurements at $180$ K, $195$ K, $210$ K and $230$ K, we find as 
expected that water crystallization becomes more and more prominent. In addition, the fraction of hexagonal ice, whose signature can be found in 
specific peaks of the structure factor, steadily increases: it is negligible within our analysis at $160$ K, around $30\%$ at $170$ K, $42\%$ at $180$ K 
and almost $62\%$ at $195$ K, and $100\%$ at $210$ K and above, with no sign of a well-defined transition between $I_c$ and $I_h$, contrary to 
what stated in [\onlinecite{murata12}]. 
(Note that the ice that forms at low temperature is generically expected to be a faulty cubic ice with stacks of hexagonal ice, {\it i.e.}, a ``stacking 
disordered'' or ``stacking fault'' ice,\cite{hansen08,moore-molinero11};  however, the  signature of $I_h$ can only be detected  in the analysis of the experimentally 
measured $S(Q)$ when its fraction is large enough and when the crystallites are big enough.)

\begin{figure}[tbp]
\centering
\includegraphics[width=\linewidth]{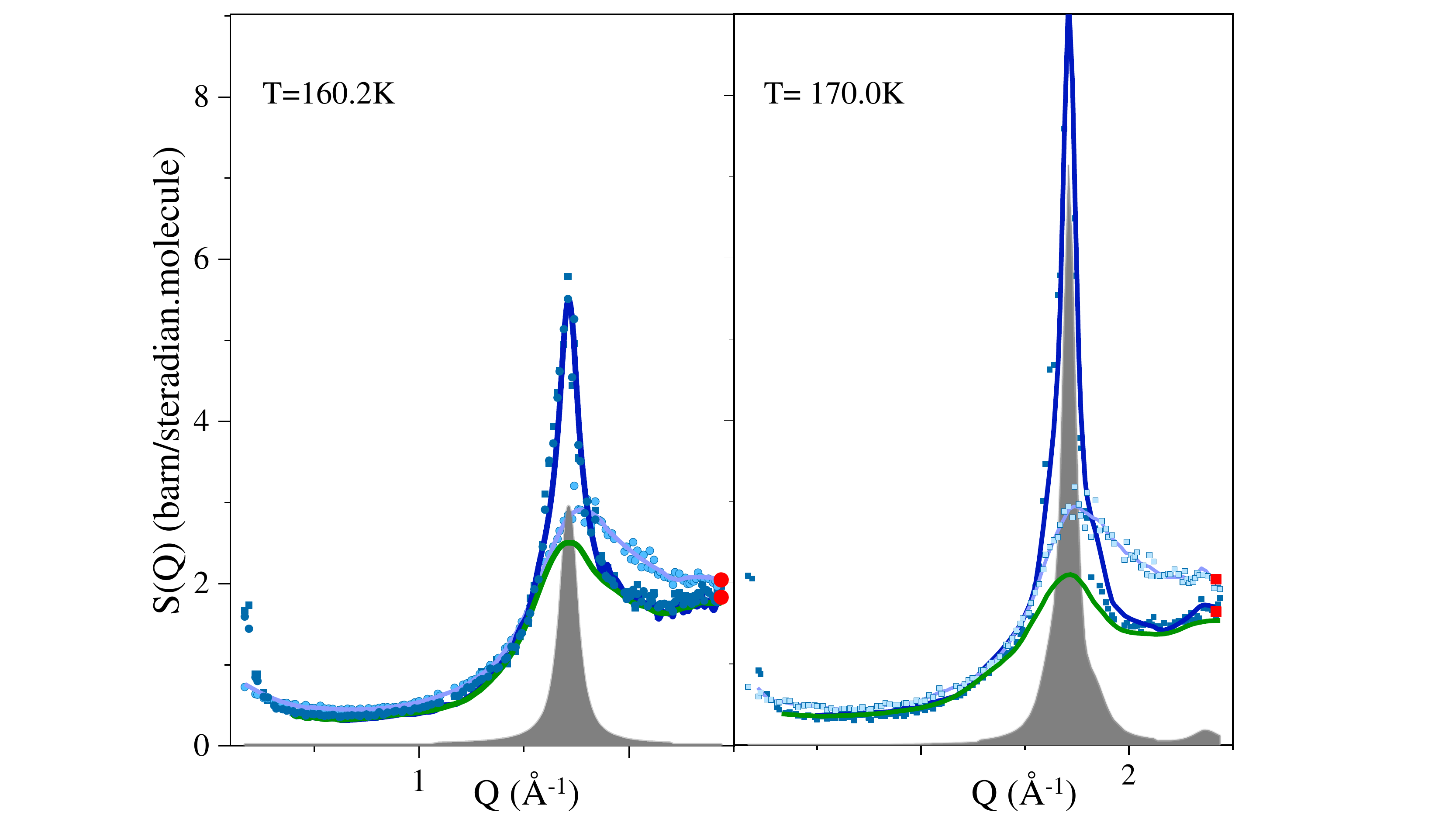}
\includegraphics[width=0.8\linewidth]{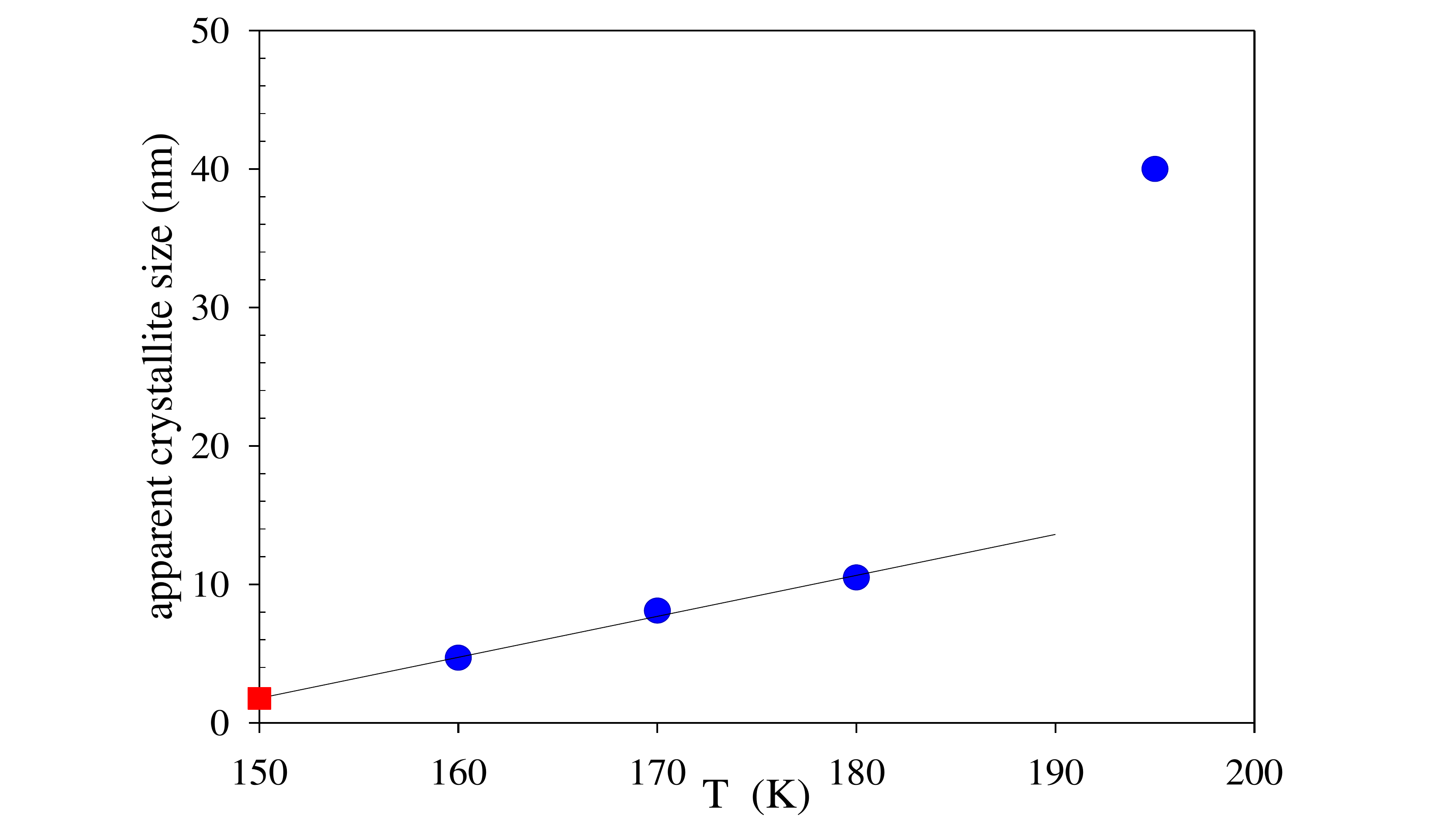}
\caption{Top: Rietveld analysis of the structure factor of the fully deuterated sample at $T=160.2$ K (left) and $T=170.0$ K (right). In both figures, 
a light (dark) color indicates the initial (final) $S(Q)$; the dark blue lines correspond to the overall fit, the shaded grey peaks are the resulting crystalline 
contribution, and the green lines are the remaining amorphous component (cut at low $Q$ to avoid showing the Porod regime). Finally the red points mark 
the initial and final coherent cross sections at high $Q$.
Bottom: Estimated apparent crystallite size as a function of the annealing temperature. The red square indicates the result of a linear extrapolation of 
the points at $160$, $170$, and $180$ K down to $150$ K.}
\label{Fig_treatment_size}
\end{figure}

{\it Crystallite size and ice fraction.} Through the analysis of the neutron scattering data we can access some of the characteristics of the ice formed in the 
transition. 

As already described, the peak that grows at $160$ K around $1.71\AA^{-1}$ is a broadened Bragg peak of cubic ice. At $170$ K the peak is 
better resolved and one sees in addition traces of sub-peaks characteristic of hexagonal ice $I_h$. The broadening of the Bragg peak is of course 
due to the limited size of the ice crystallites, and from the upper panels of Fig.~\ref{Fig_treatment_size} it is clear that  
crystallization of water leads to larger crystallites at $170$ K than at $160$ K. To get an estimate of the typical size we have proceeded to a Rietveld 
analysis\cite{rietveld} of the data at the end of the annealing process (both for the fully deuterated ${\rm C_3D_5(OD)_3+D_2O}$ and the partially deuterated 
${\rm C_3H_5(OD)_3+D_2O}$ samples, with similar outputs). The procedure is detailed in Sec. Methods and leads to an estimated apparent size of 
$4.7\pm0.3$ nm at $160$ K and of $8\pm0.5$ nm at $170$ K. The same analysis gives estimates of $10.5\pm0.6$ nm at $180$ K and $40$ nm at $195$ K: 
see Fig. \ref{Fig_treatment_size} (bottom).

Another feature which is potentially crucial for cryopreservation is the fraction of water that crystallizes into ice. One expects (see also [\onlinecite{murata12}]) 
that this fraction increases with the annealing temperature. To estimate the fraction of ice in the solution we focus on the scattered intensity at the high-$Q$ 
limit of our data, near $2.5\AA^{-1}$, whose variation reflects the decrease of water content in the solution due to ice formation (see Sec.~VIII of the SI).  We 
find that the fraction of water that has crystallized is $21\%$ at $160$ K and $39\%$ at $170$ K. 
\\

\begin{figure}[tbp]
\centering
\includegraphics[width=1.05\linewidth]{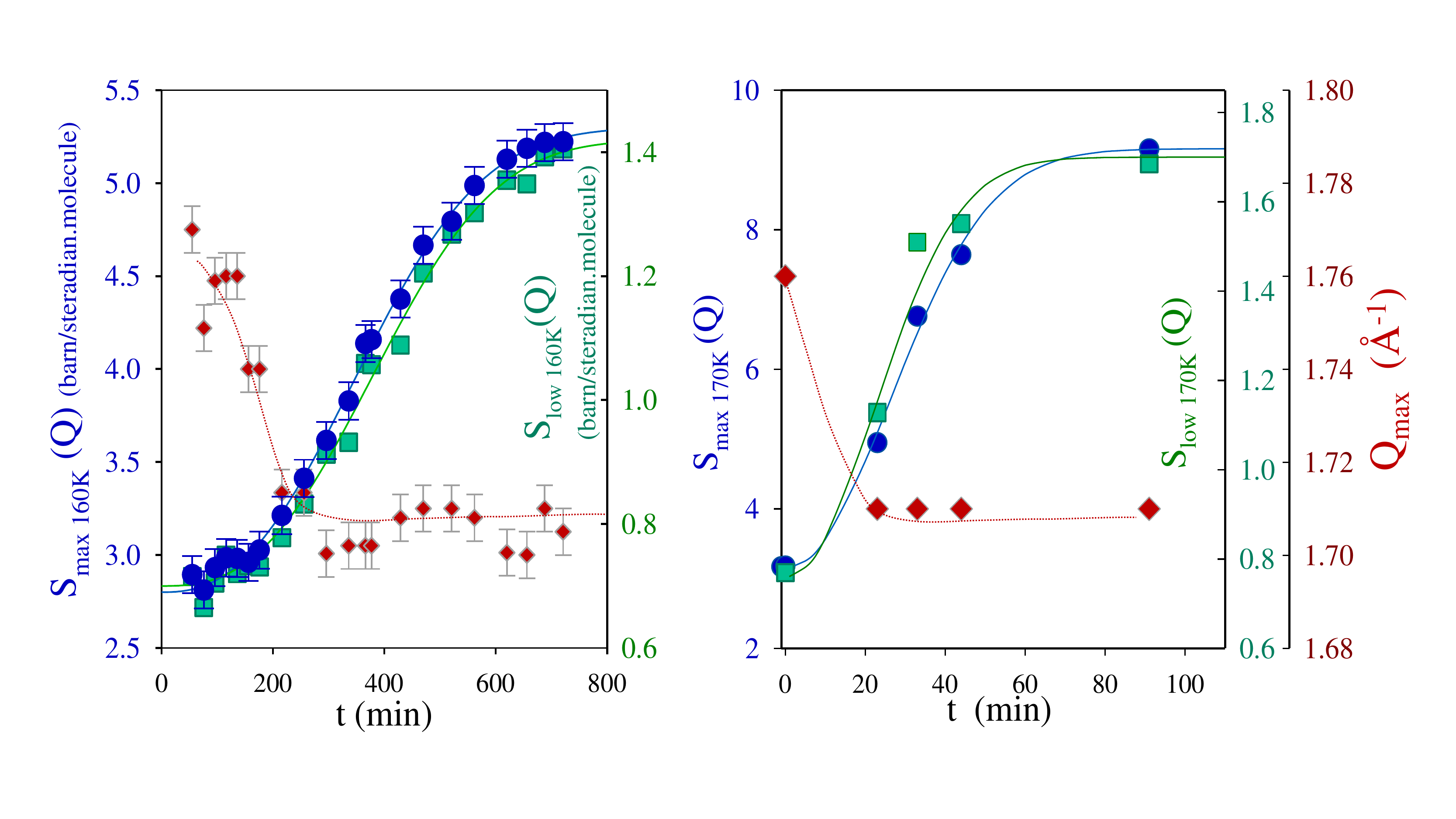}
\includegraphics[width=1\linewidth]{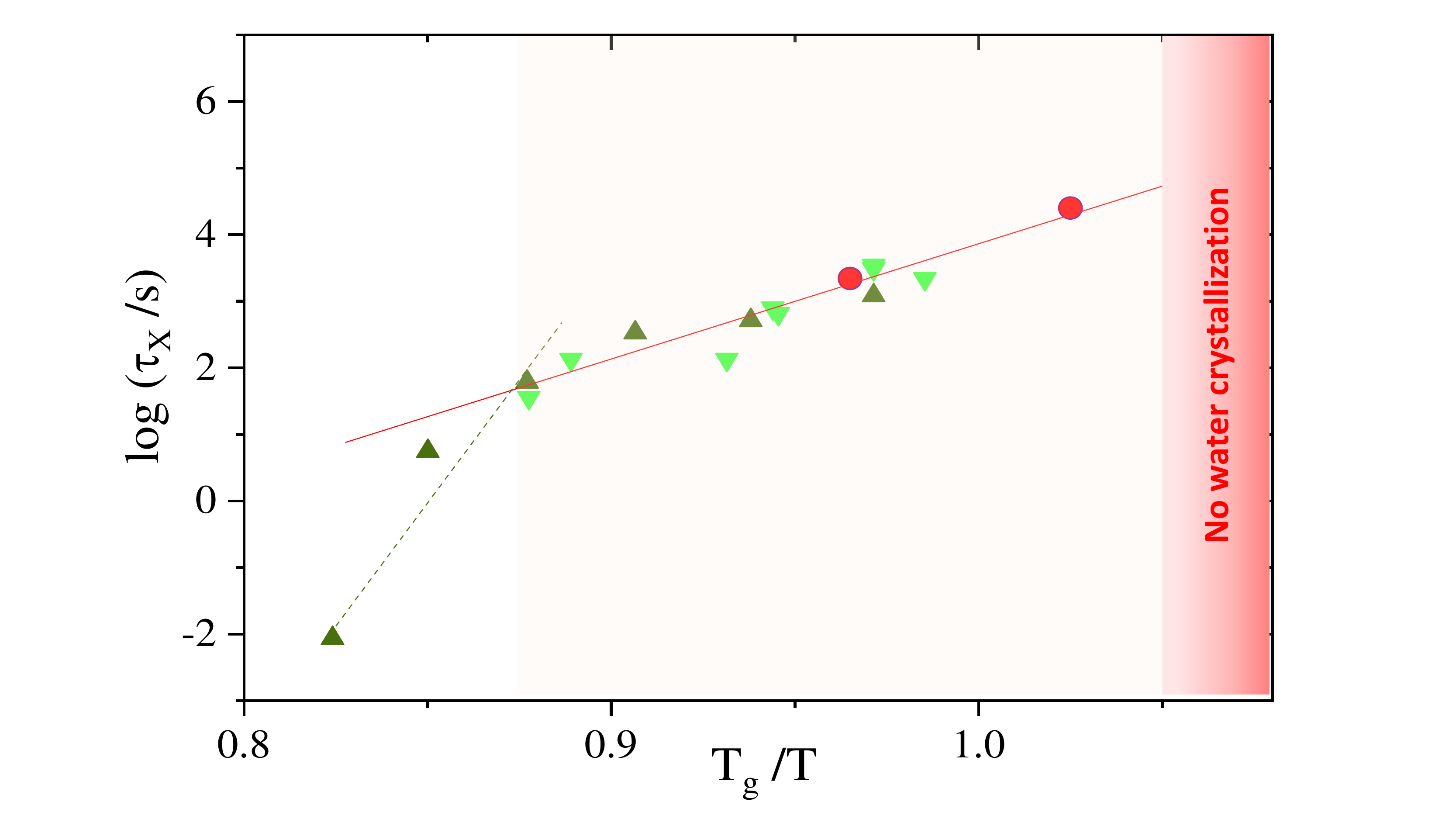}
\caption{Top: Crystallization kinetics at $T=160.2$ K (left) and $T=170$ K (right). The maximum of the structure factor $S_{\rm max}$ (blue circles) and the intensity at the 
lowest measured $Q$, $S_{\rm low}$ (green squares), are plotted versus time in minutes. The lines are the best fits to an Avrami formula (see the main text). 
We also display in both cases the evolution with time of the location $Q_{{\rm max}}$ of the main peak (red diamonds and right scale).
Bottom: $\log_{10}$ of the crystallization time to ice (in seconds) versus $T_g/T$ for the $c_g=0.178$ water-glycerol 
solution (red filled circles) and for either hyperquenched\cite{hage94} (light-green downward pointing triangles) or compressed/decompressed\cite{lin18} 
(dark-green upward pointing triangles) pure water at low temperature below $155$ K (full red line and associated symbols) where temperature is 
rescaled by the glass-transition temperature: $T_g=164.7$ K for the fully deuterated water-glycerol solution and $T_g=136$ K for pure water. We show for comparison 
higher temperature data for pure water (dashed green line and associated symbols), which clearly display a distinct behavior\cite{lin18}. 
Finally, the red region where no water crystallization takes place is estimated from the minimum crystallite size (see Sec. \ref{Discussion}).}
\label{Fig_xtal-kinetics}
\end{figure}

{\it Kinetics and timescale.} The kinetics of transformation at constant temperature is followed by monitoring several quantities: the maximum of  the peak of the 
structure factor, $S_{\rm max}$, the intensity at the lowest probed wavevector $Q=0.19\AA^{-1}$, $S_{\rm low}$, and the location of the main peak, $Q_{{\rm max}}$. 
The results for $T=160$ K and $170$ K are displayed in Fig. \ref{Fig_xtal-kinetics}. A first  observation is that ice formation appears to saturate 
after some annealing time.

A key feature is the time needed for water crystallization, which can be read from the evolution of $S_{{\rm max}}$ and $S_{{\rm low}}$. At $160$ K, crystallization 
only starts after $100$ min and takes around $800$ min to be completed, with a typical timescale $\tau_{\rm x}\sim 420$ min. The process is significantly faster at 
$170$ K because crystallization starts after only a few minutes and is completed in less than $100$ min with a typical timescale $\tau_{\rm x}\sim 37$ min. By 
assuming an Arrhenius temperature dependence\cite{hage94,molinero10,kay16}, one then obtains an estimate for the typical timescale for crystallization of 
$\tau_{\rm x}\sim 5400$ min, {\it i.e.}, almost $4$ days, at $150$ K.

From the transformation kinetics, we can also analyze the time dependence of the crystallized fraction $\Delta(t)$ through an Avrami fit\cite{avrami}, 
$\Delta(t)/\Delta(\infty)=1-\exp[K t^n]$ where $K=(1/\tau_{\rm x})^n$ depends on the temperature and on the geometry of the crystal grains and $n$ is the 
Avrami exponent that is characteristic of the growth mode\cite{papon_book}. By fitting our data, with $\Delta(t)$ obtained from either $S_{{\rm max}}$ or $S_{{\rm low}}$ 
(see Fig. \ref{Fig_xtal-kinetics}), we obtain $n\approx 2.5-2.7$ for $T=160$ K and $n\approx 2$ for $T=170$ K. Note that the trend is fully 
compatible with the measurement done at $T=195$ K for the same water-glycerol solution which gives an Avrami exponent of $n\approx 1.7$\cite{bruijn16}. 
The value of $n$ cannot be univocally interpreted, but if one assumes that the nucleation rate is constant and that growth occurs by the diffusion of water 
molecules toward the existing grains one can conclude that the geometry of the grains is spherical at $T=160$ K and elongated 
at $T=170$ K\cite{papon_book,footnote_avrami}. (The latter is compatible with the more important contribution of ice $I_h$ which tends 
to grow in a more anisotropic way than $I_c$.)

Finally, we observe that the evolution of $Q_{{\rm max}}$ (see Fig. \ref{Fig_xtal-kinetics}) is quite different than that of the crystallized fraction itself: It is much more 
rapid at the beginning, where it rather abruptly jumps from $\approx1.75$ to $1.71\AA^{-1}$, and then stays essentially constant while growth takes off. This is the 
manifestation of a first stage in the transformation that consists of nucleation of (faulty) cubic ice. Growth of the grains proceeds when this first stage 
has been in part completed.

We expect that crystallization is controlled by the diffusion of water molecules as it is for pure water\cite{hage94,hage95,kay97,molinero10,kay16}. Below the 
glass transition of the solution, the glycerol-rich matrix in which nano-segregated water domains are embedded becomes very rigid as temperature further 
decreases, so that  diffusion of water should be closer to that in a nano-confined environment than in pure water and therefore even slower than in the 
bulk. This effect may be difficult to describe in detail but it is tempting to speculate that it can be accounted at the zeroth-order level through a rescaling 
of the temperature by the appropriate glass-transition one. Accordingly, we speculate that the dependence on the annealing temperature of the 
crystallization time to ice at low enough temperature, whether in pure water, water-glycerol solutions, or more generally dilute enough aqueous 
solutions, is controlled by the dimensionless ratio $T/T_g$ where $T_g$ is the glass transition temperature of the system of interest. (This can only be valid if 
water is sufficiently nano-segregated in the solution and does indeed crystallize to ice; it is clearly not the case for a homogeneous solution with a high 
concentration of glycerol, say $c_g\gtrsim 0.28$: see the discussion below.)  
To test the idea, we have plotted in the lower panel of Fig. \ref{Fig_xtal-kinetics} our results for the fully deuterated $c_g\approx0.18$ water-glycerol solution with 
literature data on pure water at low temperature below $\sim155$ K\cite{hage94,lin18}. As can be seen from the plot, the rescaling of temperature by $T_g$ indeed 
provides a very good description. We will discuss the potential benefits of this scaling below. 
\\

{\bf Evidence against an isocompositional liquid-liquid transition}

A point of great fundamental interest raised by the results of Tanaka's group\cite{murata12,murata13} is the possibility in a window of glycerol concentration, 
roughly between $15$ and $20\%$, of an ``iso-compositional'' liquid-liquid transition triggered by a liquid-liquid transition of water itself between a low-density and a 
high-density form. Strong arguments have already been given by several groups against this 
interpretation\cite{suzuki14,popov15,feldman_more,loerting16,loerting19}, but Tanaka has reiterated his claim in a recent paper\cite{tanaka20}.

Murata and Tanaka acknowledge that water crystallization comes in the way of the putative liquid-liquid transition but they consider it as an extraneous 
phenomenon that can be subtracted. They furthermore predict that below a temperature $T_L$, which is lower than the range they studied, no crystallization 
should be seen and a pristine liquid-liquid transition could be observed. For the fully hydrogenated sample with $\sim18\%$ of glycerol, $T_L$ should be around 
$162$ K, {\it i.e.}, slightly above the calorimetric glass transition. With the temperature translation discussed above, one then expects a $T_L$ of about $166$ K for the 
fully deuterated sample. Our experiment, done with a rapid quench similar to that used in Ref. [\onlinecite{murata12}], is thus safely below the putative $T_L$ 
when annealing is considered at $T=160$ K. The outcome, detailed above, is that provided one is patient enough, water crystallization does take place at this 
temperature and explains the first-order-like transition that is observed. There is no reason to invoke an underlying liquid-liquid transition, and, even less so, an 
isocompositional one: with the fraction of crystallized water that we have estimated, the glycerol molar fraction of the remaining part of the sample is 
$21.6\%$ at $160$ K and and $24.3\%$ at $170$ K instead of the initial composition of $17.8\%$. 

Furthermore, as we discuss below in more detail, the structure of the liquid/glass found immediately after the fast quench and that of the liquid coexisting 
with ice after partial water crystallization are both dominated by a low-density amorphous (LDA) form of water.
\\

\begin{figure}[tbp]
\centering
\centerline{\includegraphics[width=1.1\linewidth]{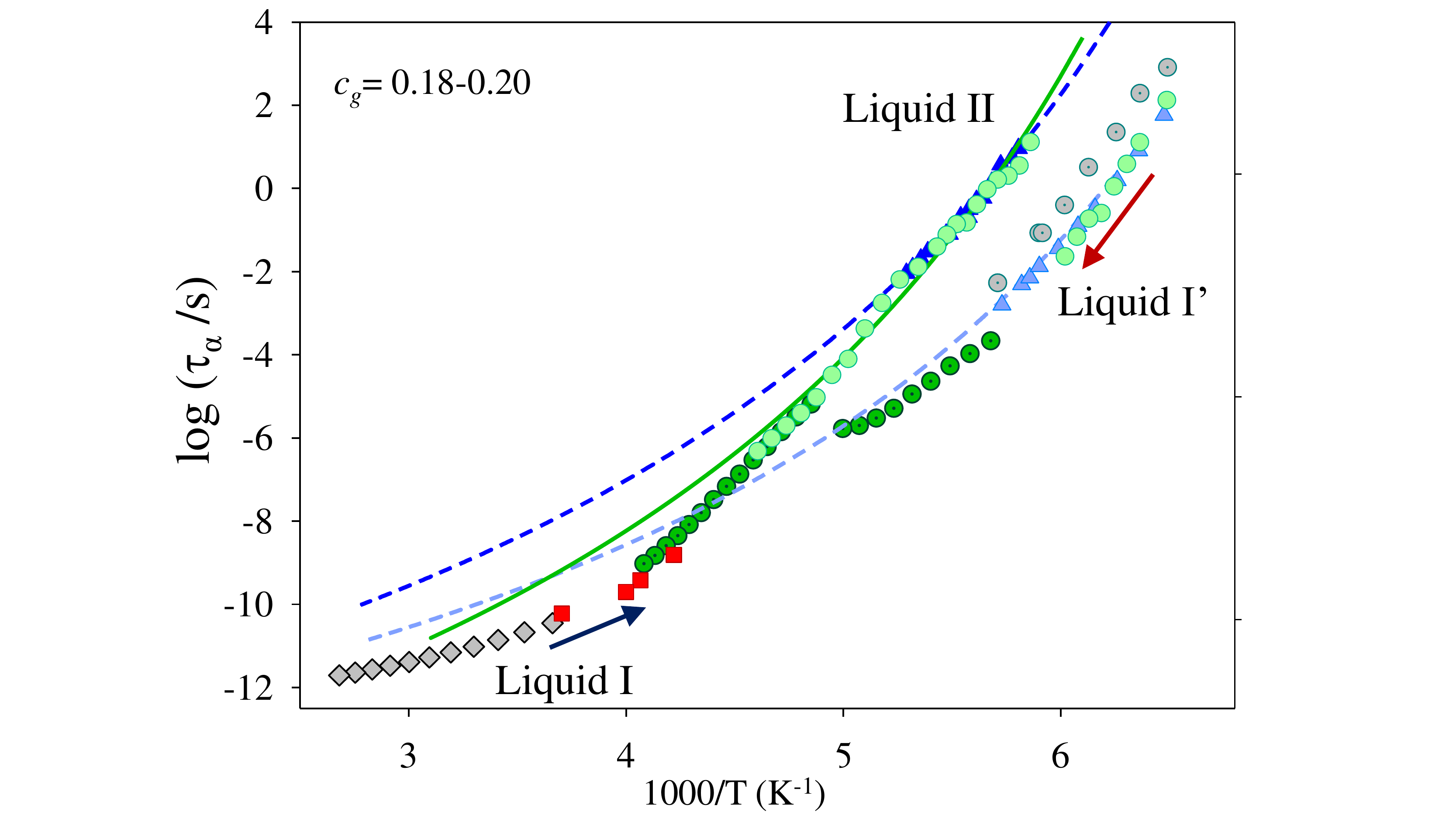}}
\centerline{\includegraphics[width=1.1\linewidth]{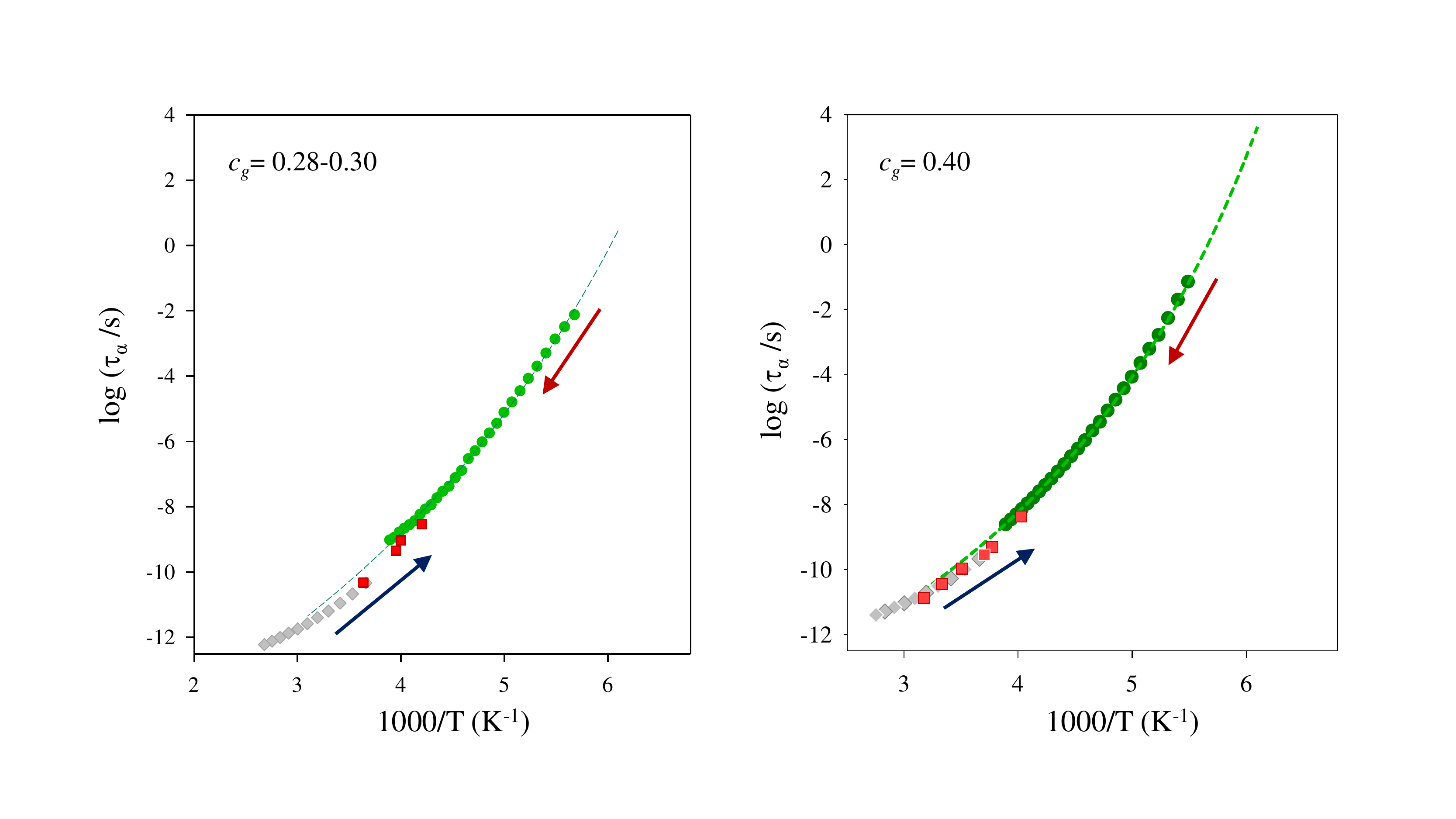}}
\caption{Arrhenius plot of the relaxation time (in seconds) and the viscosity (rescaled to have an effective time) of the glycerol-water solution for several concentrations: 
Top: $c_g=0.18$-$0.20$; Neutron Spin Echo results ($c_g=0.178$, red squares) are from the present study, dielectric data are from [\onlinecite{popov15}] 
($c_g=0.20$, green circles), [\onlinecite{murata12}] ($c_g=0.18$, blue triangles) and [\onlinecite{zhao15}] ($c_g=0.18$, grey circles), and viscosity is 
from [\onlinecite{viscosity}] ($c_g=0.18$, grey diamonds). Dark and light shades correspond to the 2 protocols in [\onlinecite{murata12,popov15}]. 
The green full line represents dielectric results at $c_g=0.4$\cite{popov15} and the dashed lines are fits to the low-temperature dielectric data\cite{murata12}. Finally, the 
arrows indicate whether the measurements are performed on cooling or on heating. As detailed in the text, liquid I denotes the stable and weakly 
supercooled liquid phase, liquid I' the out-of-equilibrium amorphous phase obtained after rapid cooling and prior to ice formation, and liquid II the liquid that remains 
when ice has formed.
Bottom left: $c_g=0.28$, except for dielectric data ($c_g=0.30$, green circles)\cite{popov15}; bottom right: $c_g=0.40$.} 
\label{Fig_dynamics}
\end{figure}

{\bf Understanding the liquid phases of the $\mathbf{18\%}$ glycerol-water solution in and out of equilibrium}

{\it Different liquid phases out of equilibrium.} The nature of the liquid phases appearing below melting at a glycerol concentration $c_g\approx0.18$ is also a vividly 
debated issue. To list the main proposals:  At low temperatures it has been suggested that the system in the glass state or before any significant annealing is a 
homogeneous solution with water in a high-density amorphous (HDA) form\cite{murata12,murata13,tanaka20} or a nano-segregated solution\cite{popov15}, possibly 
with HDA water\cite{loerting16,loerting19}; on the other hand, after heating and/or sufficient annealing, the proposals include a transformation to  a homogeneous 
solution with LDA water (called liquid II\cite{murata12,murata13,tanaka20}), a phase-separated system with coexistence of ice domains, interfacial water and a 
water-glycerol liquid mixture close to the maximally-freeze concentrated solution\cite{popov15,zhao15,loerting16,loerting19}, or a supercritical liquid fluctuating 
between low- and high-density forms of water and prone to crystallization\cite{suzuki14}. However, above all, one should stress that the phases observed in this 
temperature range are out of thermodynamic equilibrium and depend on the thermal treatment. In the following, we distinguish three liquids: 
{\it liquid I}, the equilibrium phase above the melting temperature $T_m$ and the continuously related weakly supercooled liquid phase; 
{\it liquid I'}, the (out-of-equilibrium) amorphous glass or liquid phase obtained by a fast quench of liquid I and considered prior to any significant annealing; 
and {\it ``liquid II''}, the (out-of-equilibrium) liquid that remains when the ice that has formed during the annealing is (hypothetically) removed.

{\it Liquid I' versus liquid I.}  Consider first liquid I'. As already mentioned, there is no sign of ice formation during the fast cooling process. However, 
except possibly for protocols such as hyperquenches, vapor deposition, or pressure-liquid-cooling vitrification\cite{suzuki20} which are not considered here, 
the liquid nonetheless evolves during the cooling process. 
Most notably, the nano-segregation which is already detected above melting (see the NMR results in Sec. X of the SI for $c_g=0.19$ and 
the computational modeling 
analysis for $c_g=0.25$ in [\onlinecite{soper16}]) further develops. This can be seen by comparing the structures of liquid I and liquid I' (see Fig.~S15 of the SI): 
the main peak that is typical of water correlations moves from $1.82\AA^{-1}$ at $260$ K 
to $1.75\AA^{-1}$, grows and sharpens, while there is a drop of the intensity above $2\AA^{-1}$ which, as already argued, denotes a decrease of correlation 
between water and the alkyl chains of glycerol. (Note that this is even more significant considering that while the main peak shifts to lowers $Q$'s the
 average density increases from liquid I to liquid I'\cite{protein}.) This increased nano-segregation is favored by the strong differential in mobility 
 between water and glycerol molecules in the solution. Water becomes increasingly more mobile than glycerol when temperature decreases, 
 as already observed over a limited range of temperature through a NMR study. 

The difference between liquid I and liquid I' is also illustrated by looking at their dynamics. To characterize liquid I we have carried out an experimental investigation 
by neutron spin echo (NSE) in both the stable and the weakly supercooled liquid regimes. Liquid I' has already been studied by 
broadband dielectric spectroscopy\cite{murata12,popov15}. We plot in Fig. \ref{Fig_dynamics} the relaxation time obtained from NSE and dielectric measurements, 
together with viscosity data for liquid I\cite{viscosity}, as a function of $1/T$ for $c_g\approx0.18$-$0.20$, $0.28$, and $0.40$.  One clearly sees 
that continuity of the data between liquid I at high temperature and liquid I' at low temperature is only recovered for the highest glycerol concentration shown 
(in this case the liquid can be supercooled all the way from $T_m$ to $T_g$). For $c_g\approx 0.18$-$0.20$ (top panel) the dielectric results also show 
the abrupt changeover between liquid I' and liquid II resulting from water crystallization, which we discuss below.

The additional hallmark of liquid I', besides being nano-segregated between water domains and a glycerol richer mixture, is that, contrary to what stated or 
implied before\cite{murata12,murata13,bruijn16,loerting16,loerting19}, there is no trace of water being in an HDA form. To make this clear, we compare in 
Fig. S14 of the SI the $S(Q)$ that we have obtained prior to any annealing with the neutron scattering curves of LDA and HDA for pure water\cite{koza,bellissent}. The 
main peak of liquid I' is not exactly at the same location as that of the LDA ($1.75$ versus $1.69$-$1.71\AA^{-1}$), which indicates that the hydrogen-bond network and 
local tetrahedral order are not as well formed and extended as in the LDA, but there is no resemblance whatsoever with the $S(Q)$ of HDA that peaks at $2.1\AA^{-1}$. 
(The same observation can be made for the X-ray diffraction pattern by comparing the data for liquid I' of [\onlinecite{murata12}] with those for HDA and LDA in 
[\onlinecite{urquidi03}].) Our conclusion is in line with the 
careful examination of the polarized Raman spectrum by Susuki and Mishima\cite{suzuki14}, contradicting a previous one by Murata and Tanaka\cite{murata12}, 
and with the analysis of liquid I at $238$ K just above melting by Towey et al.\cite{soper16}. Note that this conclusion is compatible with the absence of a 
thermodynamic signature for a first-order transition between LDA and HDA in compression/decompression experiments in the glass\cite{suzuki14,loerting16} if one 
takes into account that the domains of amorphous water are small (of the order of a nm): the transformation from LDA to HDA under pressure is then expected 
to be very gradual with no clear signature of a transition in thermodynamic measurements.

{\it Nature of ``liquid II''.} A final question is the nature ``liquid II''. We have shown that water crystallization takes place even below the glass transition, 
provided one waits long enough, and that ice formation saturates to a given fraction that depends on the annealing temperature 
(for a given fast-quench protocol). Unfortunately, there is no rigorous way to subtract the contribution of ice from the measured structure factor. If 
one nonetheless proceeds, as first done in [\onlinecite{murata12}] for X-ray scattering, one 
obtains the curves at $160$ K and $170$ K displayed in Fig. \ref{Fig_treatment_size} (upper panels). The remaining 
structure factor that is obtained after subtraction is distinct from that of liquid I' at the same temperature but the difference is rather subtle: one finds a small shift of the 
main peak position from $\approx1.75\AA^{-1}$ to a value around $1.71\AA^{-1}$ which roughly corresponds to the peak position of the LDA (see above). 
Popov et al.\cite{popov15} have convincingly interpreted their dielectric and DSC data as evidence for what one may call a mesoscopically phase-separated system with 
water appearing in three different configurations: (i) ice domains, (ii) hydrogen-bonded to glycerol molecules in a more-or-less water-saturated glycerol matrix, and (iii) at 
the interface between the ice domains and the matrix. It therefore appears plausible that the shift of the main peak to a lower $Q$ observed for the residual 
structure factor results from an increased contribution of a better tetrahedrally organized interfacial water as well as from the correlations between this interfacial 
water and the ice. The latter represents a cross-contribution that cannot be removed from the total spectrum, and the structure obtained by subtraction is only a 
proxy for that of a ``liquid II''.

If one now considers the evolution of liquid II with increasing temperature, one must account for the fact that more water crystallization takes place. It is then 
unlikely that the liquid could keep the same composition, be it the original one\cite{murata12,murata13,tanaka20} or the maximally-freeze concentrated 
one\cite{popov15,loerting16,loerting19}.  Actually, as seen in the upper panel of Fig. \ref{Fig_dynamics} for $c_g\approx0.18$-$0.20$, there is a clear signature of 
the sudden change between liquid I' and liquid II, a change that is a consequence of the partial crystallization of water in the solution. 
However, upon slow heating, the $T$-dependence of the dielectric relaxation time of the latter departs more and more 
from the curve of the $30$-$40\%$ concentrated mixture to finally meet that of liquid I at high enough temperature. Liquid II is thus an out-of-equilibrium phase, 
resulting from the transformation of water into ice and whose nature evolves with temperature.

\section{Discussion}
\label{Discussion}

In addition to a better understanding of the nature of the phases formed out of equilibrium by reheating rapidly quenched water-glycerol solutions and of the 
interplay between vitrification and ice formation, the present study brings some lessons for cryoprotection. First, we show that, even a few degrees below the 
calorimetric glass transition, slow water crystallization occurs in an aqueous solution with $18\%$ of glycerol; this takes place on a timescale of the order of 
$10$ hours, after which ice formation saturates. Second, we are able to estimate for this solution key parameters that control the potentially damaging effects 
of ice formation---size and shape of the ice crystallites, fraction of water converted to ice, crystallization time---as a function of the annealing temperature. 
We find by extrapolating our data that annealing the solution at a still lower temperature, say of $10$ K below the lowest one we have studied ($160$ K in 
the present case), would lead to a strong increase of the crystallization time that could reach almost $4$ days. However, such a time may not be long enough 
for long-term cryopreservation. 

This brings us to take a different angle on the question of the lower temperature limit for water crystallization. Rather than envisaging time as being the limiting 
factor, one should consider the typical size of the ice crystallites. It has been shown by a combination of experimental and simulation techniques that ice 
$I$ no longer appears when the ice clusters contain less than $90$ molecules at $150$ K (and $70$ molecules at $\sim 100$ K)\cite{moberg19}. This 
corresponds to a size (diameter) of about $1.6$-$1.7$ nm. Using a linear extrapolation of the last $3$ data points in Fig. \ref{Fig_treatment_size} 
(red line in bottom panel) gives an estimate for the crystallite size of about $1.7\pm0.1$ nm at $150$ K. This strongly suggests that ice can no longer 
form at this temperature. 

We now propose to combine this finding with the speculation discussed above that the temperature dependence of water 
crystallization in dilute aqueous solutions is controlled by the dimensionless ratio $T/T_g$ to provide a rule of thumb for estimating the lowest temperature at which 
water crystallization can happen in such an aqueous solution: multiply the $\sim150$ K estimated here by the ratio of the $T_g$ of the chosen aqueous solution and 
the $T_g$ of the fully deuterated $18\%$ glycerol-water solution, {\it i.e.}, $165$ K: $T_{{\rm lowest}}({\rm solution})\sim (150/165)\,T_g({\rm solution})$. 
Note as a cross-check of the soundness of this relation that it predicts that the lower temperature limit for ice formation in pure water is $\sim 124$ K, a temperature 
that seems indeed quite reasonable\cite{hage94,koza,kay16}.

What is the range of aqueous solutions to which this rule-of-thumb may apply? First, although we have focused on 
cryoprotectant solutions in which organic nonionic molecules such as glycerol are added to water, electrolytic solutions of water and salt which have been extensively 
studied are also concerned. Second, the solution should be dilute enough so that water at low temperature is nano-segregated and/or retains some of its bulk 
characteristics. To get a rough estimate of the upper solute concentration below which this would hold, one could (i) extrapolate the line of homogeneous nucleation of ice 
and determine when it crosses the glass transition line\cite{fahy84}, (ii) locate the crossover concentration at which the (positive) difference between the macroscopic 
density of the glass and that of the liquid above melting starts to rapidly decrease as concentration further decreases\cite{protein}, or (i) use structural indicators such that 
the presence of a rise in the low-$Q$ part of the structure factor below $0.4\AA^{-1}$ and/or of a shift of the main peak to lower $Q$'s (toward the LDA location) upon 
quenching the solution to a glass. The usefulness of the latter criterion, which requires that the main peak is sufficiently sensitive to the correlations among water 
molecules, may depend on the details of the solution: for small solute molecules such as ethylene glycol other contributions involving the solute may be involved, 
which obscures the behavior of water, and for ionic solutions, one has to take into account the specific organization of water in solvation shells around the ions. 
For water-glycerol solutions, the above criteria all seem to restrict the range to $c_g\lesssim 0.21-0.22$: see Fig. S1 and Sec. I of the SI. For the salt solution of LiCl 
in water, the upper bound would be a molar concentration of around $12\%$ from criterion (i)\cite{angell_aqueous} and $16\%$ from criterion (ii)\cite{LiCl_chieux}. 
The rule-of-thumb for, say, a $10\%$ solution then predicts  that no ice formation is possible below $T_{{\rm lowest}}\approx (150/165)139\approx126$ K. 
More investigations are certainly necessary to refine the prediction and the range of solutions to which it applies.

As far as cryopreservation is concerned, preventing ice formation is one prerequisite. One should also be able to heat back the sample at room temperature without 
too much damage caused by ice formation and thawing during the heating process. Considering what we have found in the present study, this seems hard to avoid 
with a simple glycerol-water solution in the ``marginal'' concentration range. Other chemicals such as ice recrystallization inhibitors would then be 
required\cite{ice_inhibitors1,ice_inhibitors2}. The knowledge of the crystallites size and shape and of the fraction of ice, as provided in the present study, are 
then potentially crucial to design the appropriate cryoprotectant mixtures.

\section{Methods}

{\bf Samples.} 
Thanks to the neutron scattering method, the contribution of partial structure factors can be selectively probed through specific H/D substitutions. The isotopic 
compositions were chosen in such a way that there was no exchange between H and D during the different thermal treatments nor any uncertainties in the analysis 
of the scattering data.  The sample compositions used for elastic neutron scattering are summarized in Table I of the SI , and detailed below in the sections 
about DSC and neutron spin echo. The sample concentration $c_g$ is expressed as the number of glycerol molecules divided by the total number of molecules. 
The samples were purchased from Sigma-Aldrich, Eurisotop and Cambridge Isotope Laboratory and used without any further purification. The deuteration rate is 
$99.9\%$ for ${\rm D_2O}$,  $99.9\%$ for fully deuterated glycerol, and above $98\%$ for partly deuterated glycerol.

{\bf Polarized Neutron Scattering experiments.}
The neutron scattering experiments were performed on the D7 spectrometer at the Institut Laue Langevin in Grenoble (France). The main advantage 
of this instrument is its ability to separate coherent scattering (containing the structural information) from incoherent scattering by using a technique of 
longitudinal neutron polarization analysis\cite{D7}. The incoherent scattering can be used as an internal calibration. In this way, one directly obtains the 
coherent scattering signal in absolute units [${\rm barn/(steradian.molecule)}$] and one avoids ``vanadium calibration'' of the instrument with all its uncertainties. 
We considered one glycerol concentration, $c_g\approx 0.18$, but several H/D substitutions, with ${\rm D_2O}$ or ${\rm H_2O}$ and fully [${\rm C_3D_5(OD)_3}$] 
or partially deuterated [${\rm C_3H_5(OD)_3}$ and ${\rm C_3D_5(OH)_3}$] glycerol, were used to independently focus on the contribution of water, of atoms 
involved in hydrogen-bond network, and of the alkyl chains. We verified  for each of the mixtures that the measured $S(Q)$ at high $Q$ in liquid I' prior to any 
crystallization is fully compatible with the calculated total coherent scattering cross section per steradian and per molecule. The former values were used as a 
basis to estimate the amount of crystallized water and the glycerol mole fraction of the remaining solution without ice (see Sec.~VIII of the SI).

Measurements were performed from $130$ K to $300$ K in an Orange cryostat by using an alumina annular cell of $0.2$ mm thickness and an external diameter of 
$19.5$ mm in order to maximize the transmission. A wavelength of $4.8 \AA$ was chosen to measure the structure factor in the $Q$-range 
$0.1\AA^{-1} < Q < 2.4 \AA^{-1}$. The raw data  obtained on D7, {\it i.e.}, measurements of the non-spin-flip scattered intensity and the spin-flip one, 
were corrected by using the standard reduction software developed at the ILL (https://www.ill.eu/users/instruments/instruments-list/d7/characteristics). The data are 
first normalized by  the monitor counts. The background, measured from the scattering of the empty cell and a black sample (Cadmium) with the same geometry, 
was subtracted after taking into account the appropriate self-absorption corrections ({\it i.e.}, the measured transmission of the sample). The flipping ratio was measured 
from a quartz rod of the same diameter, and the detectors' efficiency was corrected with a vanadium sample. The incoherent and coherent signals were then calculated 
from the non-spin-flip and spin-flip scattered intensities. Absorption and multiple scattering were kept negligible by ensuring with a high transmission that only few \% 
of the neutrons were scattered and by choosing cylindrical sample geometry (of only $0.2$mm thickness). Thanks to the method, inelastic events included in the 
incoherent scattering were properly subtracted.

Complementary structural investigations were performed at the Laboratoire L\'eon Brillouin (LLB) on the instrument G44 
(See http://www-llb.cea.fr/fr-en/pdf/g44-llb.pdf) which provides an increased $Q$-range at the expense of a lower flux and no neutron polarization. Experiments 
then take longer. A wavelength of $2.89 \AA$ and a cylindrical vanadium cell of $6$ mm diameter were used. Results are illustrated for 
$c_g=0.28$ in Sec. XI of the SI.

{\bf Rietveld analysis.}
The Rietveld method allows an easy modelling of the peak shape, width and intensity of diffraction patterns in relation with the atomic structure of crystalline 
phases and their relative contribution\cite{rietveld}. The Rietveld analysis of the polarized neutron scattering data was performed with the FullProf program 
(see https://www.ill.eu/sites/fullprof/ and [\onlinecite{juan}]). The refinement of the scattered intensities was carried out as a function of the scattering angle  $2\theta$, 
which is the experimental quantity for a monochromatic diffractometer as D7 (the wave vector $Q$ is then defined as $Q = 4\pi \sin\theta/\lambda $ 
where the wavelength $\lambda$ is $4.8\AA$), by accounting for the instrumental resolution as detailed in [\onlinecite{resolution}]. The peak intensity 
and shape were fitted by taking into account an averaged crystal structure based on the hexagonal $I_h$ and cubic $I_c$ crystalline structures of 
pure ${\rm D_2O}$\cite{iceD2O_structure} and their possible combination. The sharpening of the peak with increasing annealing time and/or  
temperature provides a measure of the average (isotropic) crystallite size. Finally, the difference between the experimental data and the 
pattern calculated from the structural model corresponds to the averaged residual amorphous part present in the sample. In summary, this analysis gives 
access to the proportion of hexagonal and cubic ice, to the average size of the crystallites, and to the remaining amorphous or liquid contribution.  

{\bf Neutron (Resonance) Spin Echo (NSE, NRSE).}
Neutron spin echo is a powerful quasi-elastic neutron scattering technique to study the dynamical properties of the mixtures in a wide momentum-transfer range 
and in the range of a few picoseconds to tens of nanoseconds\cite{NSE}. Neutron spin echo is based on the neutron spin property,  
{\it i.e.}, its spin rotation, which encodes the energy transfer occurring during the scattering process. Before and after the process, a magnetic field 
is applied which generates the precession of the neutron and only depends on the velocity difference of each neutron, irrespective of the initial velocity. As a result, 
this difference is independent of the chosen wavelength, which means that NSE can use a wide distribution of wavelength while keeping its resolution and boosting 
the signal intensity. The outcome is the normalized intermediate scattering function $F(Q,t)$; it contains the contributions from both coherent and incoherent functions. 
However, in a fully deuterated sample the coherent part can be easily extracted, which gives access to the collective component of the correlation function.

The instrument MUSES at the LLB (Saclay, France) combines conventional and Resonance Neutron Spin Echo (NRSE)\cite{muses}. The conventional NSE 
spectrometer is used for measurements at small (so-called Fourier) times ($t < 200$ ps) and the NRSE option gives access to measurements at longer times 
($200$ ps$< t < 2000$ ps). The experiments are carried out with an energy resolution of $0.3$ $\mu$eV with an incident neutron spectrum of 
$\delta\lambda/\lambda=0.15$ bandwidth. We studied fully deuterated samples with several glycerol mole fraction, $c_g = 0.178, 0.28, 0.40$, to cover the whole 
domain II of the phase diagram in Fig. S1 of the SI. Typical curves are shown in Fig. S11 of the SI for $c_g = 0.178$ and several temperatures 
from $200$ to $280$ K at the wave vector $Q=1.9\AA^{-1}$ corresponding to the maximum of the structure factor $S(Q)$ in this range of temperature.

{\bf Differential Scanning Calorimetry (DSC).} 
DSC is a thermal analysis that measures the heat flow associated with materials transitions as a function of the time and temperature, in a controlled atmosphere. 
In the present case, DSC  with a cooling rate and a heating rate of $10$ K/min was used to determine the temperature of the glass transition of the 
deuterated samples as well as their melting/freezing phase transitions plotted in Fig. S1 of the SI. No signature of crystallization was 
observed when cooling at $10$ K/min (crystallization only occurred on heating); however, crystallization was observed upon cooling at a rate of $2$ K/min. 
DSC scans are shown in Sec.~VII of the SI. 

{\bf High-resolution and pulsed-field gradient $^1$H NMR (NMR, PFG-NMR).}
The NMR sample was prepared with fully protonated glycerol and water at a molar fraction $c_g=0.19$. The mixture was immediately sealed in a glass tube 
of $4$ mm diameter. NMR spectra were recorded with a Bruker Avance spectrometer at $9.4$ T ($^1$H resonance frequency: $400.13$ MHz) and a standard 
dual broadband $5$ mm probehead equipped with a gradient coil. A Bruker temperature controller unit using evaporated liquid nitrogen flow allows experiments 
between $180$ K and $330$ K with an accuracy and stability of $\pm 2$ K. Temperature calibration was performed with a standard 
methanol reference tube. Measurement of the area of the different signals gives access to the ratio of mobile and immobile species in the sample, hence to the 
fraction of crystallized water\cite{NMR1}.
 
Self-diffusion coefficients were measured by PFG-NMR and stimulated echo sequence\cite{NMR2}. The maximum magnitude of the pulsed field gradient 
was $60$ G.cm$^{-1}$, the diffusion delay $\Delta$ was adjusted between $50$ ms and $1000$ ms, and the gradient pulse length $\delta$ was set between 
$1$ ms and $5$ ms depending on the diffusion coefficient. The self-diffusion coefficients were determined from the classical Stejskal-Tanner equation\cite{NMR3}, 
$\ln(I/I_0) = -D G^2 \gamma^2 \delta^2(\Delta-\delta/3)$, 
where $G$ is the magnitude of the two applied gradient pulses, $\gamma$ is the gyromagnetic ratio of the nucleus under study, and $I$ and $I_0$ are the 
integrated intensities of the signal obtained with and without gradient pulses, respectively. Here, we used $16$ equally spaced gradient steps for each 
experiment. Data acquisition and treatment were performed with the Bruker Topspin software.
   
{\bf Thermal treatments.}
For the structural studies we quenched the liquid samples in liquid nitrogen from $295$ K down to $\sim 80$-$90$ K. This roughly corresponds to a cooling rate of 
$70$-$130$ K/min.  The sample temperature was fully stabilized at $90$ K and then taken at $130$ K at which measurements were performed to 
characterize the glass phase. The samples were further heated to the chosen annealing temperature close to the glass temperature, {\it i.e.}, $160$ K and $170$ K, 
and we followed the evolution of the structure for long annealing times until the signal was completely stabilized. To give an idea, this took around $13$ hours at 
$160$ K. Finally we further heated the sample.
We also studied another protocol corresponding to a slow cooling of $3$-$6$ K/min down to the glass at $130$ K. We then heated the sample to the temperature 
at which the measurements were performed: see Sec.~IV of the SI.

\begin{acknowledgements}
We thank I. Popov et Y. Feldman for fruitful discussions and access to their dielectric data, A. Wildes who was the local contact on D7 at the ILL, M. Hartmann for 
discussion on the structural data, F. Legendre for technical support at the LLB (Muses), and M. Bombled for helping on the DSC experiments at the LLB. 
We also thank the eRMN team of ICMMO (University Paris-Saclay) for providing access to the NMR facility and J. Texeira for useful exchanges and a careful reading of the manuscript.
\\

{\bf Author contributions}

C. A.-S. and G. T. conceived the research and wrote the manuscript, C. A.-S., G. T., and F. C.  performed the neutron scattering experiment on D7 
for which C. A.-S. analyzed the data; O. O. carried out the DSC measurements and C. A.-S. analyzed them; F. P. contributed to the diffraction experiments on G44 
and to the Rietveld analysis of the structural data; P. J. performed and treated the NMR experiments; S. L., O. O., and C. A.-S. conducted and analyzed 
the NSE measurements.
\\

{\bf Additional information}

The authors declare no competing financial interests.

\end{acknowledgements}

\pagebreak

\appendix{\bf{SUPPLEMENTARY INFORMATION}}

\begin{figure}[b]
\centerline{
\includegraphics[width=1.2\linewidth]{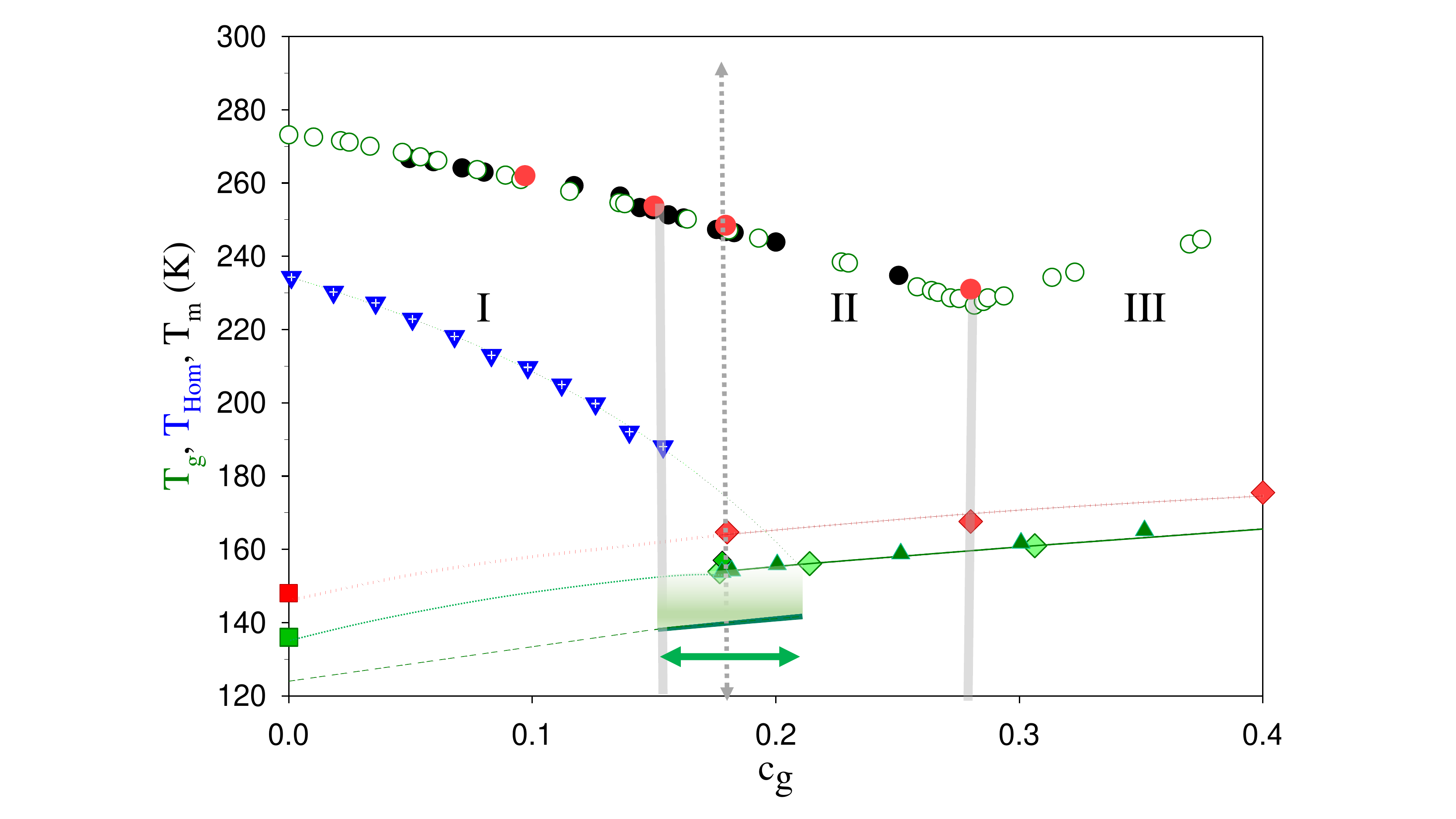}}
\caption{Phase diagram of the water-glycerol mixture in the temperature versus glycerol (molar) concentration plane. We show the melting line 
(open\cite{lane25} and filled\cite{murata12} black circles) as well as the glass transition line when no crystallization takes place on cooling (dark\cite{murata12} and 
light\cite{inaba07} green symbols) for fully hydrogenated samples. 
The filled red symbols are our results for fully deuterated samples (whose characteristic temperatures are moved up by $3$-$4$ K at melting and $8$-$10$ K at the glass transition\cite{footnote_quantum}). The lines and the extrapolations toward $c_g=0$ (pure water) are only guides for the eye. 
As described in the main text, the phenomenology below the melting line is usually divided in $3$ regions denoted I, II, and III\cite{popov15} 
(see also~[\onlinecite{nakagawa19, zhao15}]). In region I water crystallization occurs on cooling (even in liquid nitrogen). The intermediate region II, or even a part of it, is 
the one of interest here, and we mark by a blue vertical dotted line the concentration which we focus on. We also display as downward pointing 
blue triangles the temperatures of homogeneous nucleation of ice in the solution\cite{broto76,kanno05} and show as a dotted line a fit and an extrapolation at 
higher concentrations where it then crosses the glass transition line around $c_g\approx 0.21-0.22$. The region between the latter concentration  
and the boundary of region I ($c_g\approx 0.15$), which is indicated by the large double-headed arrow, is where we expect nano-crystallization of water to take place in 
the glassy state (obtained after a rapid quench in liquid nitrogen). The full green line in this region is our prediction for the temperature below which ice formation is no 
longer possible and the extrapolation down to $c_g=0$ is only indicative.}
\label{Fig_messy_phasediagram}
\end{figure}

\section{Temperature-concentration phase diagram of water-glycerol solutions}

We show in Fig.~\ref{Fig_messy_phasediagram} the phase diagram of water-glycerol solutions where 
three different ranges of glycerol concentration can be distinguished below the melting line. The range of prime interest for the present study 
is that in which water is nano-segregated and is part of region II. In region I, water crystallizes upon even rapid cooling (hyperquenches and alternative techniques 
may still prevent this to happen, as for pure water) and one can define a line of homogeneous nucleation where micrometer size crystallites of ice are 
formed when cooling (e.g., by using an emulsification method)\cite{broto76,kanno05}: the experimental points are displayed as downward pointing (blue) triangles 
in Fig.~\ref{Fig_messy_phasediagram} and the data can be extrapolated as shown by the dotted blue line. The extrapolation crosses the glass transition line 
around $c_g\approx 0.21-0.22$. This value can tentatively be taken as the upper limit of the region where nano-segregation is possible. We suggest that a 
signature of nano-segregation of water is when the macroscopic density of the solution increases upon quenching, as in a standard molecular liquid,  while the 
main peak of the structure factor shifts to smaller $Q$, as is characteristic of bulk water. Unfortunately, we have not covered in detail the range between $c_g=0.15$ 
and $c_g=0.22$. We find that the phenomenon does appear for $c_g\approx 0.18$ but is no longer present for $c_g=0.28$ (see Sec.~XI below). An additional 
criterion is provided by looking at the (positive) difference in macroscopic density between the glass and the high-temperature liquid above melting: This difference 
is rather constant at high concentration but starts to rapidly decrease around $c_g\sim0.21$\cite{protein}. More work is nonetheless needed to more precisely specify 
the domain where nano-segregation takes place after rapid cooling to a glass and to extend the study to other aqueous solutions.

\section{Details on the characteristics of the samples.}

We provide in table~\ref{table1} some details on the samples: isotopic composition, glycerol molar fraction $c_g$, mass of an equivalent molecule, where what 
is called a ``molecule'' represents a fictitious molecular unit made of $(1-c_g)$ molecules of water and $c_g$ molecules of glycerol, incoherent and coherent 
scattering cross-sections, transmission of the sample on the D7 diffractometer, and finally coherent scattering cross-section per steradian and per molecule.

\begin{figure}[tbp]
\centering
\includegraphics[width=\linewidth]{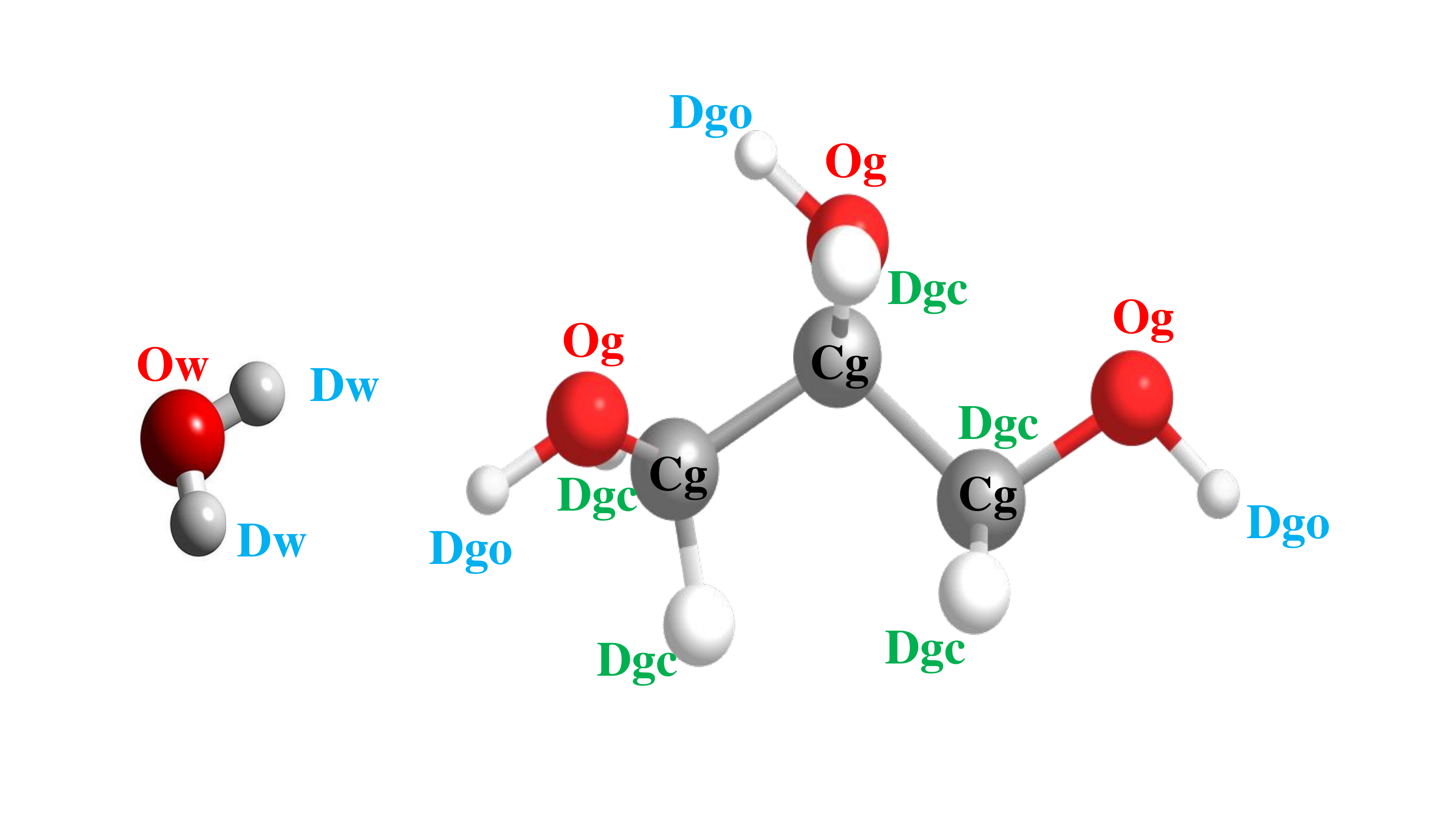}
\caption{Sketch of the molecules with their atoms labeled according to Table \ref{table1}.}
\label{Fig_molecules}
 \end{figure}

In Fig.~\ref{Fig_molecules} we sketch the molecules of the water/glycerol solutions with their atoms labeled according to Table~\ref{table1}.
\\

\begin{table}[t]
\resizebox{\columnwidth}{!}{
\begin{centering}
\begin{tabular}{cl  c c c c c c c}
\hline\\
sample & ${\rm mass}$ & glycerol & pseudo-molecule  & $\sigma_{{\rm inc}}$ & $\sigma_{{\rm coh}}$ 
& transmission & $\sigma_{{\rm coh}}/(4\pi)$ &\\
  &  /g & molar fraction & mass/g & /barn & /barn 
&  &/ster.molec &\\[2 ex]
\hline\\
${\rm C_3 \textcolor{green}{D_{gC}}_5(\textcolor{red}{O_g}\textcolor{blue}{D_{gO}})_3\textcolor{blue}{D_{w2}}\textcolor{red}{O_w}}$ 
& 1.4031 & 0.178 & 34.31 & 6.2942 & 25.705 & 0.9 & 2.046 &\\[2ex]
\hline\\
${\rm C_3 H_5(\textcolor{red}{O_g}\textcolor{blue}{D_{gO}})_3\textcolor{blue}{D_{w2}}\textcolor{red}{O_w}}$  
& 1.4966 & 0.178 & 33.37 & 76.602 & 22.392 & 0.81 & 1.777 &\\[2ex]
\hline\\ 
${\rm C_3 \textcolor{green}{D_{gC}}_5(\textcolor{red}{O_g}H)_3 H_2\textcolor{red}{O_w}}$  
& 1.4743 & 0.177 & 32.05 & 176.603 & 17.481 & 0.67 & 1.391 &\\[2ex]
\hline
\end{tabular}
\end{centering}
}
\caption{Numerical values characterizing the mixtures of glycerol and water for neutron scattering. The sample 
${\rm C_3 H_5(O_g D_{gO})_3 D_{w2}O_w}$ is only $98\%$ deuterated.}
\label{table1}
\end{table}

\section{Absence of crystallization upon rapid cooling.}

\begin{figure}[tbp]
\centering
\includegraphics[width=\linewidth]{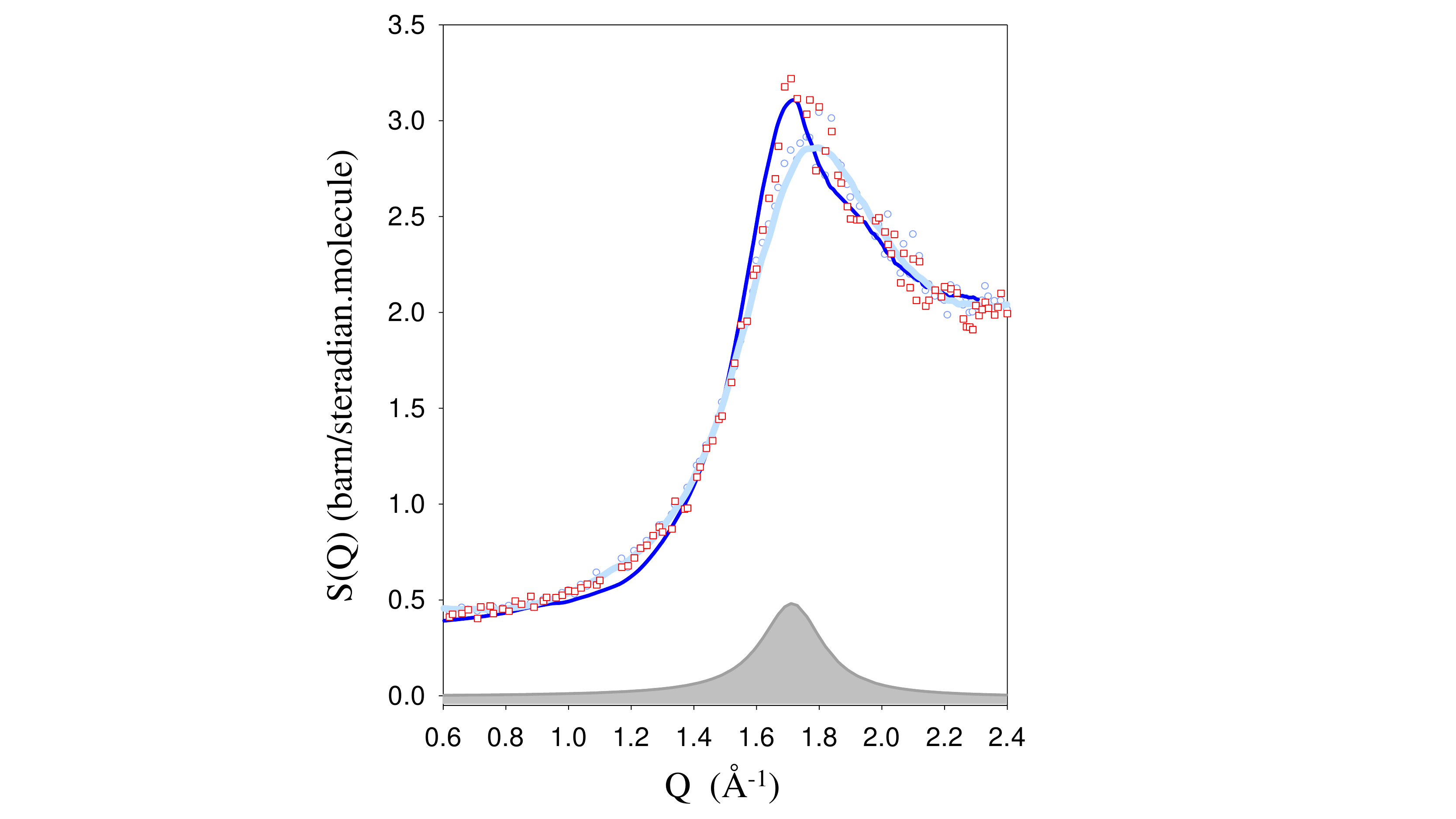}
\caption{Structure factor $S(Q)$ of ${\rm C_3D_5(OD)_3+D_2O}$ for $c_g=0.178$ and $T=160.2$ K obtained by a 
fast quench in liquid nitrogen. The light blue symbols and the associated interpolating curve correspond to the data  prior to any annealing and the red symbols to 
the data after an annealing of $216$min. The dark blue curve is the best Rietveld fit of the data prior to annealing when assuming the presence of ice crystallites of 
$1.7$nm (whose diffraction spectrum is shown by the gray shaded area) on top of an amorphous background. One can see that the agreement is very poor and that 
the putative system with $1.7$nm crystallites rather corresponds to the structure after $216$min, where water crystallization has indeed started.}
\label{Fig_fast_noxtal}
 \end{figure}

Evidence that the glassy samples with $c_g\approx 0.18$ obtained through a rapid enough cooling showed no sign of crystallization is as follows:

The structure factor $S(Q)$ after a fast quench in liquid nitrogen and prior to annealing shows no detectable signs of crystallization: the peak is at a $Q\approx1.75\AA^{-1}$, somewhat higher than that of ice (and of the LDA), and when we observe crystallization upon isothermal annealing, the first stage 
is precisely a shift of the peak maximum (see Fig. 3 of the main text); the amplitude of the main peak is also significantly lower than when crystallization proceeds. 
This can be quantified by resorting to a Rietveld analysis (details on the analysis are given in Sec. Methods of the main text) of the structure factor. Assuming that small 
ice crystallites of $1.7$nm of diameter are present (this is roughly the lower limit size for which ice can form, see Sec. Discussion of the main text) on top of an amorphous 
background, we look for the best fit to the experimental data. The result is shown in Fig. \ref{Fig_fast_noxtal}. One can see that the fitted peak is too high and shifted 
to the left compared to the actual data, even for such a small typical crystallite size. As also displayed in the figure, we find that the reconstructed structure factor with 
$1.7$nm size crystallites rather corresponds to the experimental data after an annealing of $216$min, i.e., when crystallization has indeed started in the sample.

A second argument against the presence of ice in rapidly cooled samples comes from the DSC measurements (see Sec. VII below). When cooled at a fast 
rate of $10$ K/min, still slower than the quench in liquid nitrogen that we use for the study of the structure, no sign of crystallization is detected in the DSC scans on cooling. 
On the other hand, crystallization is observed by using a slower rate of $2$ K/min.

Fast enough cooling ($10$ K/min or faster) therefore seems to completely suppress crystallization for the concentration $c_g\approx 0.18$.
\\

\section{Influence of the thermal treatment on the structural data.}

\begin{figure}[tbp]
\centering
\centerline{\includegraphics[width=1.1\linewidth]{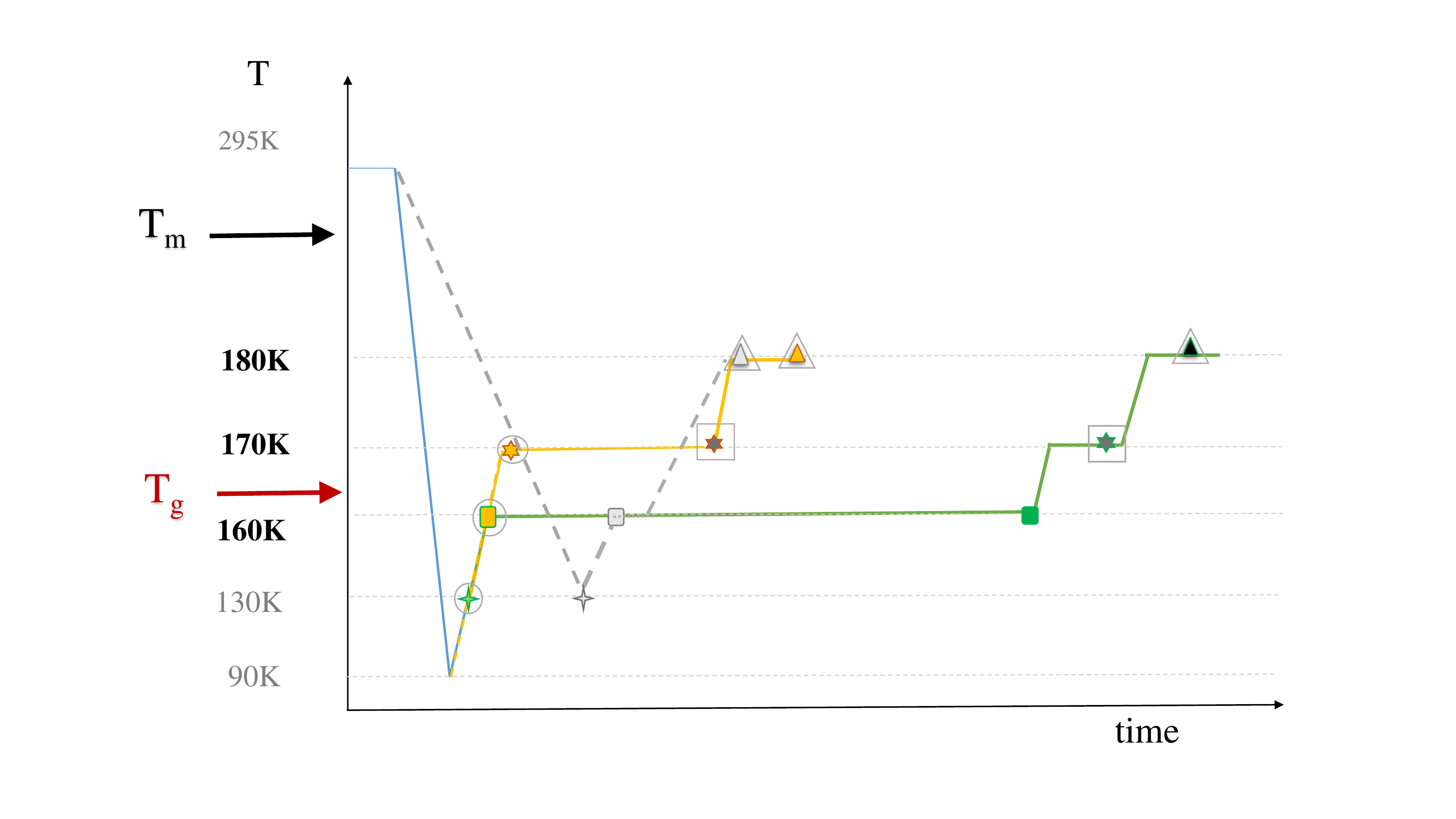}}
\caption{Sketch of the thermal treatments used in this study. The full blue line and the gray dashed line respectively represent the fast cooling process and the 
slow cooling one. The symbols indicate some of the points at which structural measurements were taken: The open squares correspond to 
Fig. \ref{Fig_S(Q)_170-180K}(left), the triangles to Fig. \ref{Fig_S(Q)_170-180K}(right), and the circles to Fig. 8 of the main text (liquid I'). The yellow and green 
horizontal line indicate the annealing stages.}
\label{Fig_Thermal_treatment}
 \end{figure}

For the neutron scattering study we have considered two cooling protocols, the fast quench in liquid nitrogen already discussed and a slow cooling at $3$-$6$ K/min down 
to $130$ K ($6$ K/min at the beginning, then $3$ K/min below $180$ K). The slow cooling is followed by heating to the chosen annealing temperature. The various 
thermal treatments are sketched in Fig.~\ref{Fig_Thermal_treatment}.

\begin{figure}[tbp]
\centering
\centerline{\includegraphics[width=1.15\linewidth]{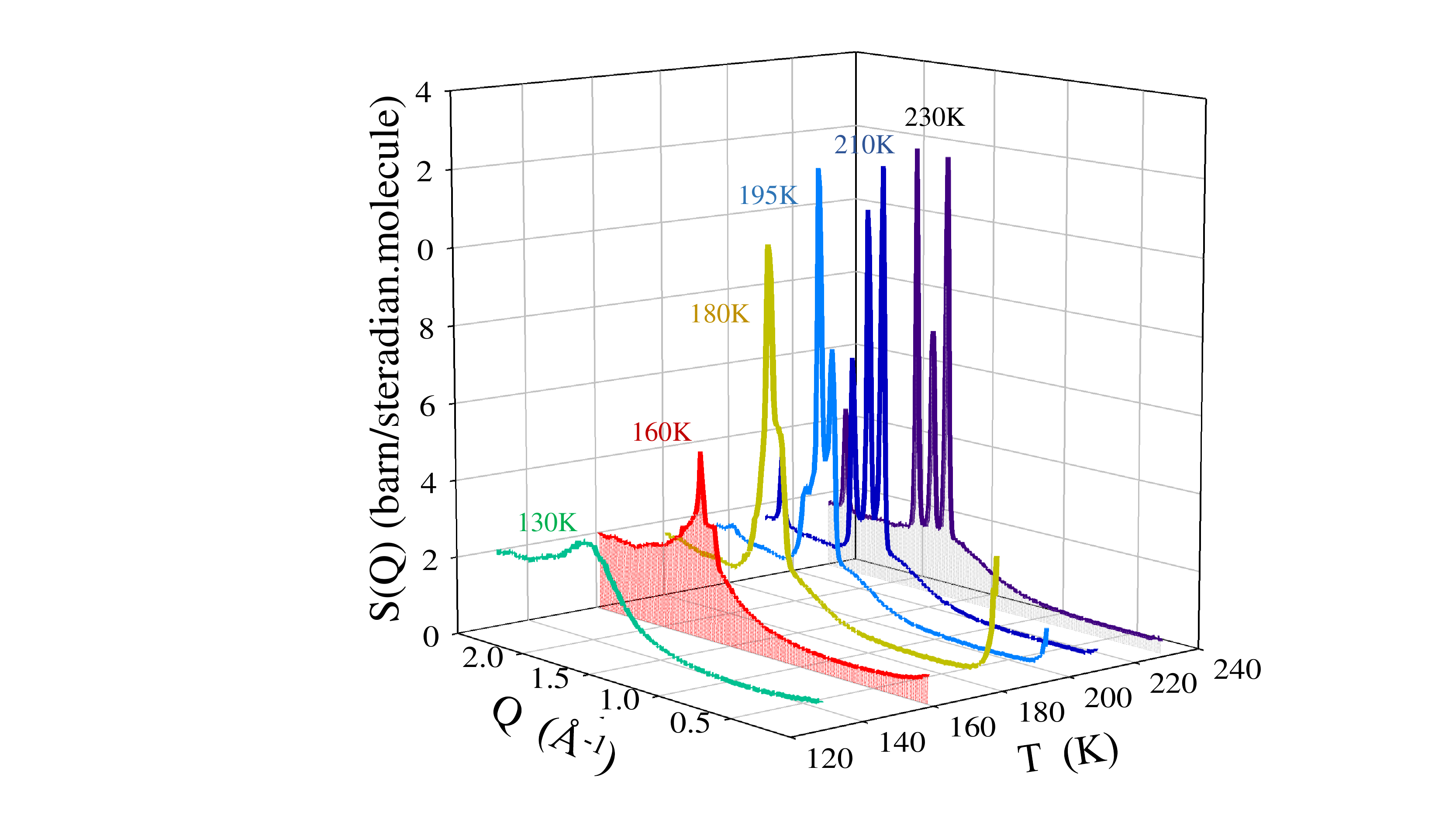}}
\caption{Structure factor $S(Q)$ of ${\rm C_3D_5(OD)_3+D_2O}$ for $c_g=0.178$ obtained by slow cooling of the sample at $3$-$6$ K/min: it is plotted for 
increasing temperatures from $130$ K to $230$ K.}
\label{Fig_3Dplot_slow}
\end{figure}

The results at $160$ K depend on the protocol. We saw that for the fast cooling in liquid nitrogen, crystallization is a slow process taking place over several hours. 
Alternately, for the slow-cooling protocol, crystallization is much faster. At low temperature, $T=130$ K, we observe a similar glass structure as for the fast 
quench, but as soon as one heats the sample, water partly crystallizes: see Fig.~\ref{Fig_3Dplot_slow}. Furthermore, whereas we do not detect within 
our experimental resolution and our analysis hexagonal ice $I_h$ after the fast quench, it appears to form about $16\%$ of the total ice after the slow 
cooling. Interestingly, the apparent grain size is similar in the two cases ($\sim 4$-$4.5$ nm for slow cooling) but the fraction of water that has crystallized is much 
smaller after a slow cooling: only $7.5\%$ to be compared to the $21\%$ in the other protocol. 

On the other hand, when the samples are heated to $180$ K, the structure factor is the same, whatever the protocol: see the right panel of Fig.~\ref{Fig_S(Q)_170-180K} below.

Fig.~\ref{Fig_3Dplot_slow} illustrates two features of cold crystallization in this aqueous solution. First, the structure evolves with increasing 
temperature from one dominated by features of the cubic ice with a peak around $1.71 \AA^{-1}$ to a mixture of cubic and hexagonal ice and finally 
to a structure dominated by hexagonal ice with its distinctive triplet of main peaks and an additional peak at $2.38 \AA^{-1}$. The change appears rather gradual, at variance 
with the suggestion that cubic ice gives way to hexagonal ice as soon as the water grains reach a critical size of about $10$-$15$nm\cite{johari05}. Second, the dramatic 
upswing at the lowest probed values of $Q$ which is observed at $180$ K is characteristic of interface formation due to many crystallites (as more than $50\%$ of 
water has crystallized in $\sim 10$nm grains: see below and main text). At lower temperatures, this effect is less visible because the fraction of ice and the number 
of grains is smaller. At higher temperatures, the size of the ice crystallites becomes too large for the upswing to appear in our experimental window: the whole Porod 
behavior moves to lower values of $Q$ to which we did not have access in the D7 experiment.

\begin{figure}[tbp]
\centering
\includegraphics[width=\linewidth]{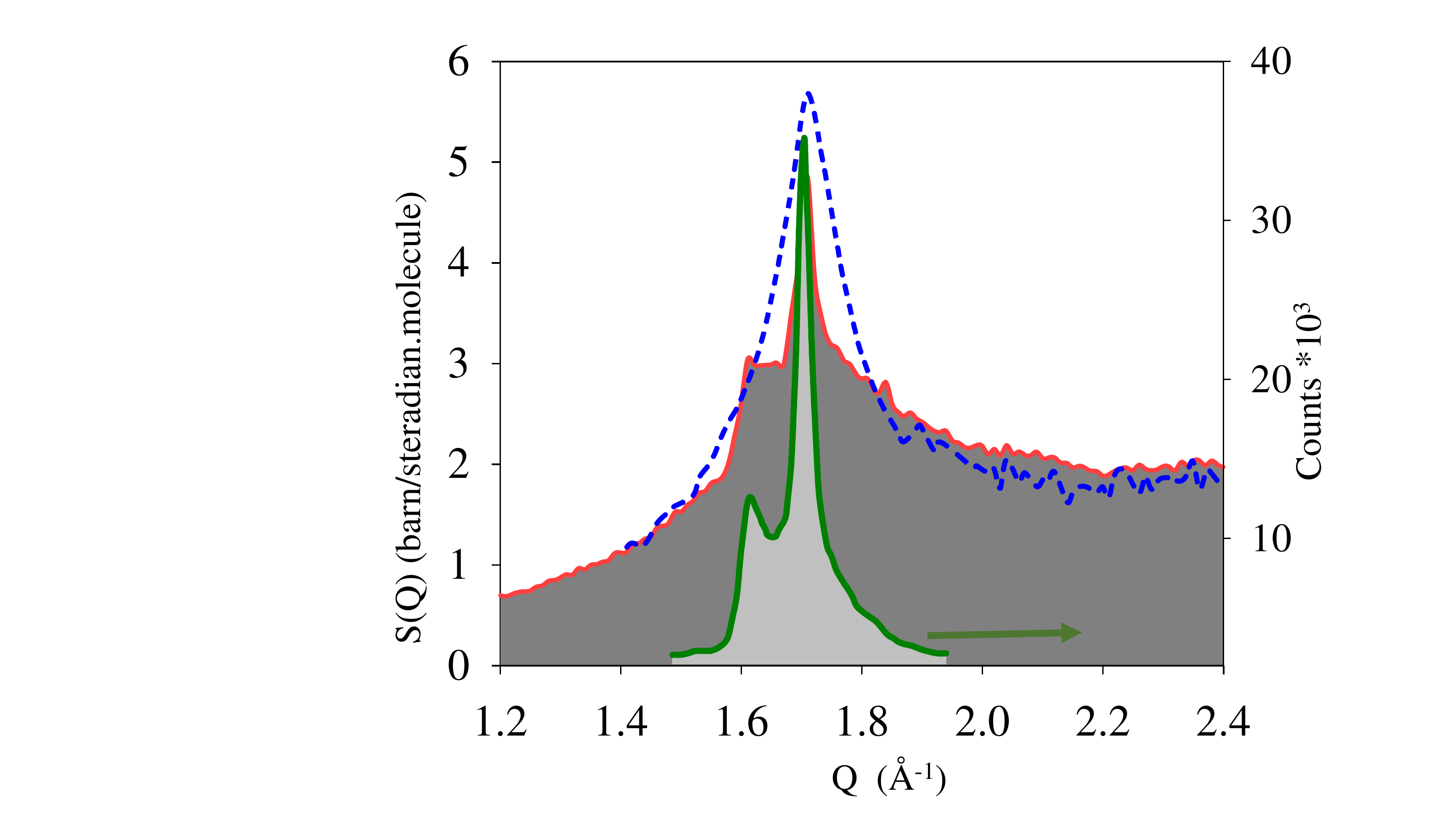}
\caption{Comparison of the structure factor $S(Q)$ of ${\rm C_3D_5(OD)_3+D_2O}$ for $c_g=0.178$ and $T=160.2$ K obtained either by a 
fast quench in liquid nitrogen at the end of the isothermal annealing (dotted blue line) or a slow cooling (red curve and dark shaded area) with that of the stacking 
disordered ice obtained from a different protocol and reported in [\onlinecite{hansen08}] (green curve and light shaded area).}
\label{Fig_slow_ice}
 \end{figure}

Finally, in Fig.~\ref{Fig_slow_ice} we compare the structure factor of ${\rm C_3D_5(OD)_3+D_2O}$ for $c_g=0.178$ and $T=160$ K obtained either 
by a fast quench in liquid nitrogen (at the end of the isothermal annealing) or by a slow cooling (see also Fig. \ref{Fig_3Dplot_slow}) with that of the stacking disordered 
ice obtained at $147$ K from a different experimental protocol starting from ice V\cite{hansen08}. One can see that there is a clear resemblance in the region 
of the main peaks between the $S(Q)$ from slow cooling and the stacking disordered ice, the latter being a cubic ice structure disordered by hexagonal 
stacking faults. There is of course an important amorphous background in the $S(Q)$ of the solution, and the Rietveld analysis that we have performed 
captures in a crude way this mixture of cubic and hexagonal symmetries. On the other hand, one can check that the $S(Q)$ 
obtained after a fast quench and prior to any annealing is very different from the two others, in accordance with what already discussed above and in the main text.
\\

\section{Robustness of the structure at $170$ and $180$ K versus the annealing protocol.}

\begin{figure}[tbp]
\centering
\centerline{\includegraphics[width=1.05\linewidth]{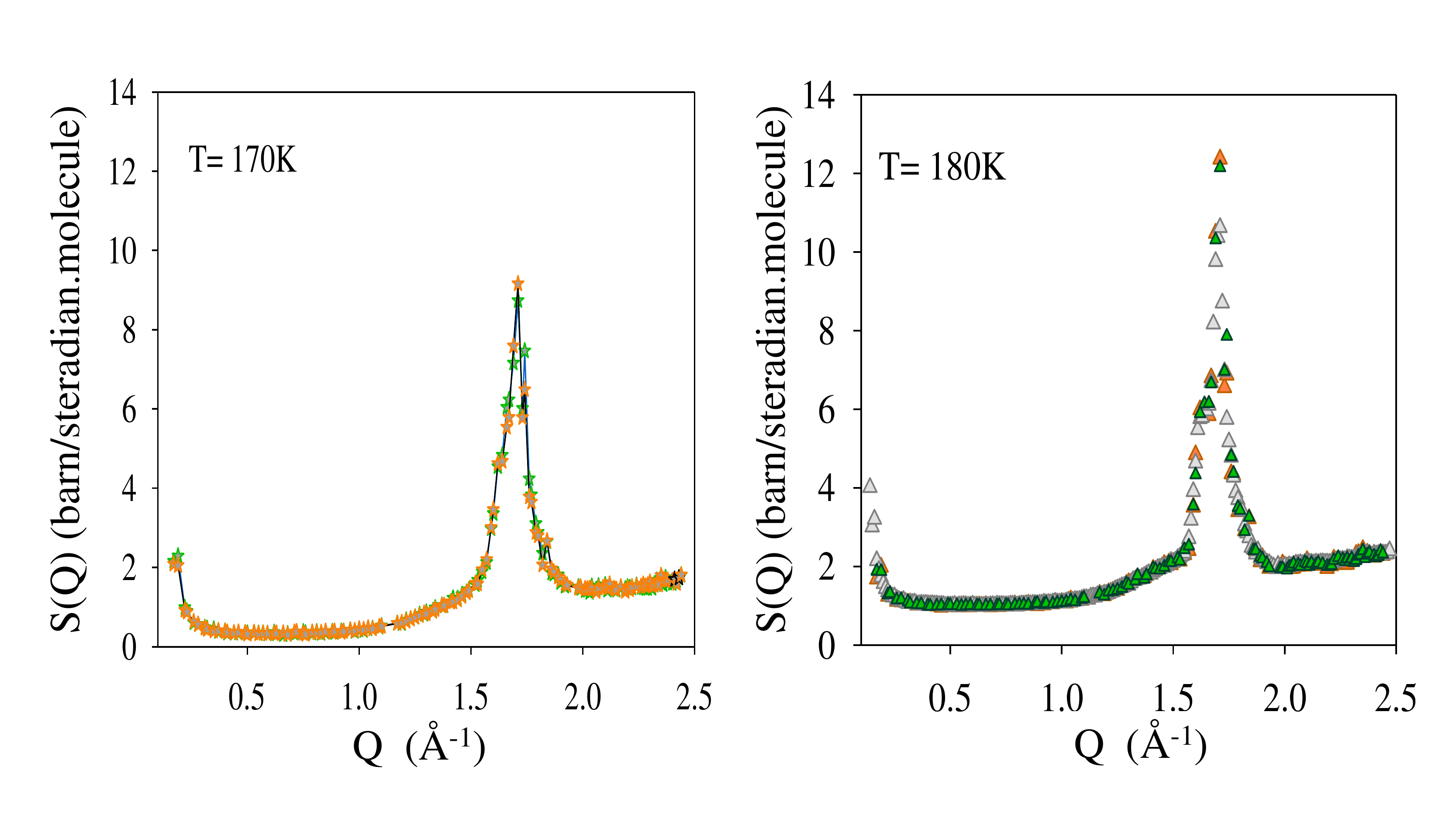}}
\caption{Structure factor $S(Q)$ of ${\rm C_3D_5(OD)_3+D_2O}$ for $c_g=0.178$ obtained through different thermal treatments. Left, $T=170$ K: Fast quench 
in liquid nitrogen, followed by heating and full annealing at $170$ K (orange stars); fast quench in liquid nitrogen, followed by heating and full annealing at $160$ K, 
then heating at $170.3$ K (green stars). Right, $T=180$ K: Fast quench in liquid nitrogen, followed by heating and full annealing at $170$ K, then heating at $180.9$ K 
(orange triangles); fast quench in liquid nitrogen, followed by heating and full annealing at $160$ K, then heating at $170.3$ K and finally at $180.6 $ K (green triangles); 
in this case we also display the result of the slow cooling in the cryostat to $130$ K, then heating at $160$ K, and finally at $180$ K (gray triangles).}
\label{Fig_S(Q)_170-180K}
\end{figure}

We illustrate here that the states obtained by a fast quench in liquid nitrogen followed by heating and annealing do not depend on the details of 
the thermal treatment: The structure factors at $170$ K are obtained either after a full annealing directly at $170$ K or after a full annealing at $160$ K followed  
by heating at $170.3$ K; similarly, the structure factors at $180$ K are obtained either after a full annealing at $170$ K followed by heating at $180.9$ K or after a 
full annealing at $160$ K, followed by heating at $170.3$ K and finally at $180.6$ K. The resulting phases with partially crystallized water are very stable 
(albeit actually only metastable) and no longer evolve with time: this is shown for $T=170$ K and $T=180$ K in Fig.~\ref{Fig_S(Q)_170-180K}.
\\

\section{Analysis of the prefactors of the low-$Q$ Porod law}

\begin{figure}[tbp]
\centering
\centerline{\includegraphics[width=1.1\linewidth]{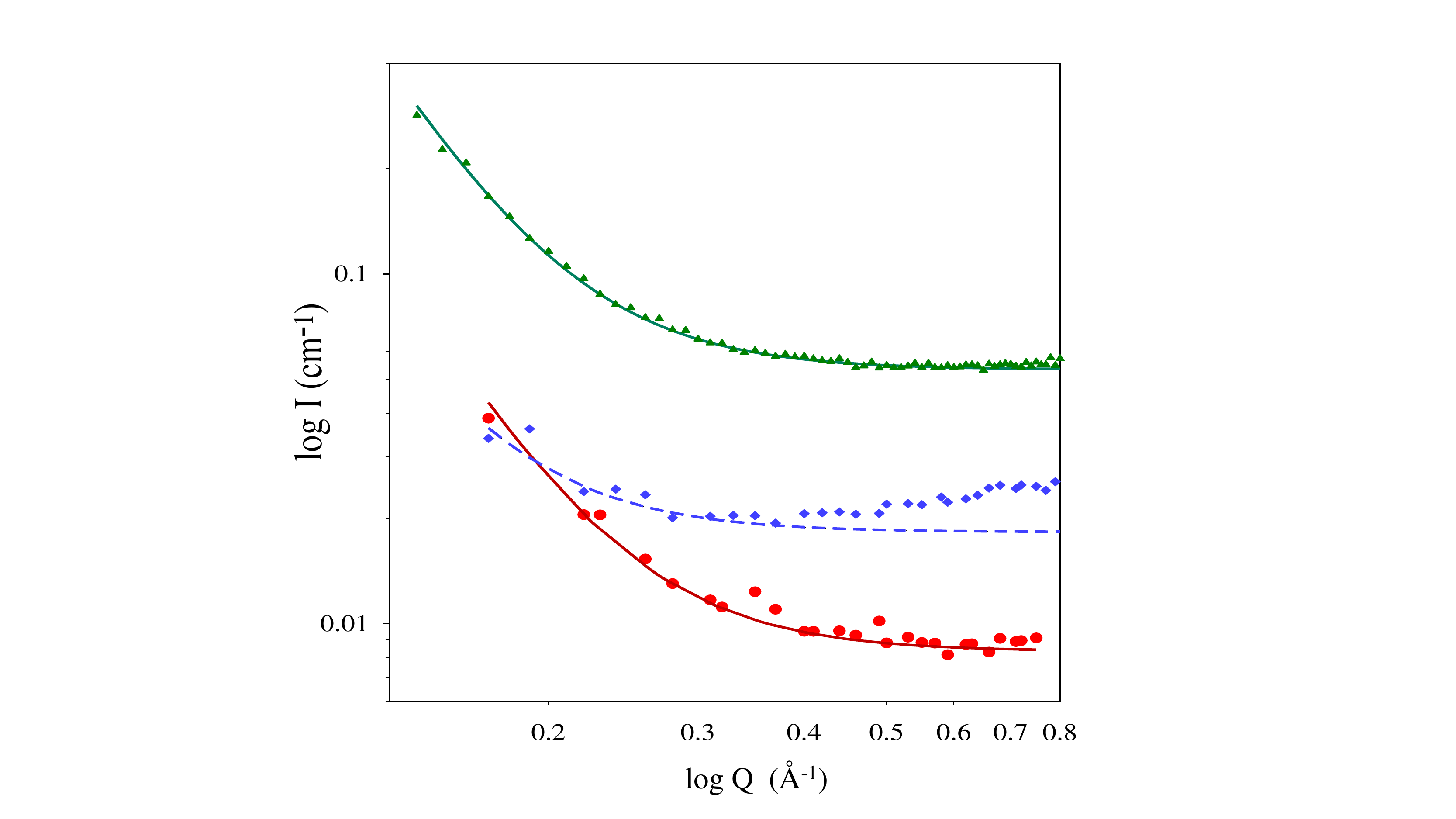}}
\caption{Log-log plot of the scattered intensity at low $Q$ for the three different deuterations of the glycerol-water solution with $c_g=0.178$ at 
$160$ K at the end of the annealing when water has crystallized: same data and color code as in Fig. 1 of the main text. The lines are the best fits to Porod's law, 
$I(Q)=aQ^{-4}+b$. The fit to the ${\rm C_3H_5(OD)_3+D_2O}$ data (dashed blue line and symbols) is only indicative as over the range of probed wavevectors 
the Porod regime does not emerge distinctly enough.}
\label{Fig_Porod}
 \end{figure}

We consider the amplitude of the $Q^{-4}$ behavior in the low-$Q$ region of the scattered intensity of the fully deuterated and partially 
deuterated samples after crystallization at $160$ K: see Fig. 1 of the main text. Porod's prediction states that in the presence of an interface 
between two phases the scattered intensity at small wavevectors goes as $I(Q)\sim a Q^{-4}$ with 
\begin{equation}
a=2\pi \left (\frac SV\right )K^2
\end{equation}
where $S/V$ is the area of the interface per unit volume and $K^2$ is the contrast between the two phases. We then fit our data for each sample 
at low $Q$ to a functional form  $a Q^{-4}+b$ with $a$ and $b$ adjustable parameters. The resulting fits are shown in Fig.~\ref{Fig_Porod}. As can be seen,  
the fits are good except for the partly deuterated sample ${\rm C_3H_5(OD)_3+D_2O}$ for which the Porod behavior is not well pronounced over the probed 
range of wavevectors due to the presence of additional contributions near and above $0.3\AA^{-1}$. 

We would like to rationalize the trend observed between the three different deuterations. However, even when looking at the ratios between the 
amplitudes $a$, there are still too many unknown parameters. We therefore make some rather crude assumptions to obtain a qualitative or semi-quantitative 
answer. First, we consider that one of the phase is a mixture of glycerol and water (a proxy for liquid II) with the glycerol concentration $c'_g\approx 0.216$ 
determined above for the fully deuterated sample (and we take the same value for the other samples) and the other phase is made of ice crystallites. 
With this assumption, the ratio of the contrasts between two samples denoted $1$ and $2$ can be obtained as
\begin{equation}
\frac{K_1}{K_2}=\left [\frac{n_{ps}(c'_g)[c'_g(\Sigma b)_{g1}+(1-c'_g)(\Sigma b)_{w1}]-n_w (\Sigma b)_{w1}}
{n_{ps}(c'_g)[c'_g(\Sigma b)_{g2}+(1-c'_g)(\Sigma b)_{w2}]- n_w (\Sigma b)_{w2}}\right ]^2 \, ,
\end{equation}
where $(\Sigma b)_{g\alpha}$ and $(\Sigma b)_{w\alpha}$, with $\alpha=1,2$, are the sums of all coherent scattering lengths for glycerol and water in sample 
$\alpha$; $n_w$ is the number density of cubic ice and $n_{ps}(c'_g)$ is the number density of pseudo-molecules comprising $c'_g$ molecules of glycerol and 
$(1-c'_g)$ molecules of water. The latter is further approximated by parametrizing the data for the mass density experimentally determined at $77$ K 
in [\onlinecite{density}] which we divide by the mass of the fully hydrogenated pseudo-molecule at the appropriate concentration: in the range of interest, a 
quadratic fit to the resulting number density is very good. Finally, we assume that the interface area $S/V$ is roughly the same for the three samples, so that 
the ratio of amplitudes $a_1/a_2$ is given by the ratio of contrasts $K_1/K_2$.

With the above simplifying hypotheses, we obtain a ratio between ${\rm C_3D_5(OH)_3+H_2O}$ (green curve in Fig.~\ref{Fig_Porod}) and 
${\rm C_3D_5(OD)_3+D_2O}$ (red curve in Fig.~\ref{Fig_Porod}) of about $3.8$ while the corresponding empirically determined one is $3.3$; on the other hand, 
the predicted ratio between ${\rm C_3D_5(OH)_3+H_2O}$ (green curve in Fig.~\ref{Fig_Porod}) and ${\rm C_3H_5(OD)_3+D_2O}$ (blue curve in Fig.~\ref{Fig_Porod})  
is $11.4$ while the corresponding empirically determined one is $6.4$. One can see that there is a semi-quantitative agreement between calculated and 
observed ratios with a correct prediction of the trend between samples. (As could be anticipated, a much larger discrepancy is found when 
${\rm C_3H_5(OD)_3+D_2O}$ is involved.) To go further in the analysis and obtain an estimate of a residual glycerol concentration that could account 
for the potential presence of interfacial water, it would be necessary to carry out systematic experiments with smaller wavevectors by Small-Angle Neutron Scattering.
\\

\section{DSC measurements.}

We have used two DSC $Q$-$100$ from TA instruments, depending on the temperature range explored. One of them was equipped with a compressor and able 
to go down to $193$ K, and the other was equipped with a liquid nitrogen cooling system and able to reach temperatures as low as $140$ K. 

\begin{figure}[tbp]
\centering
\centerline{\includegraphics[width=1.2\linewidth]{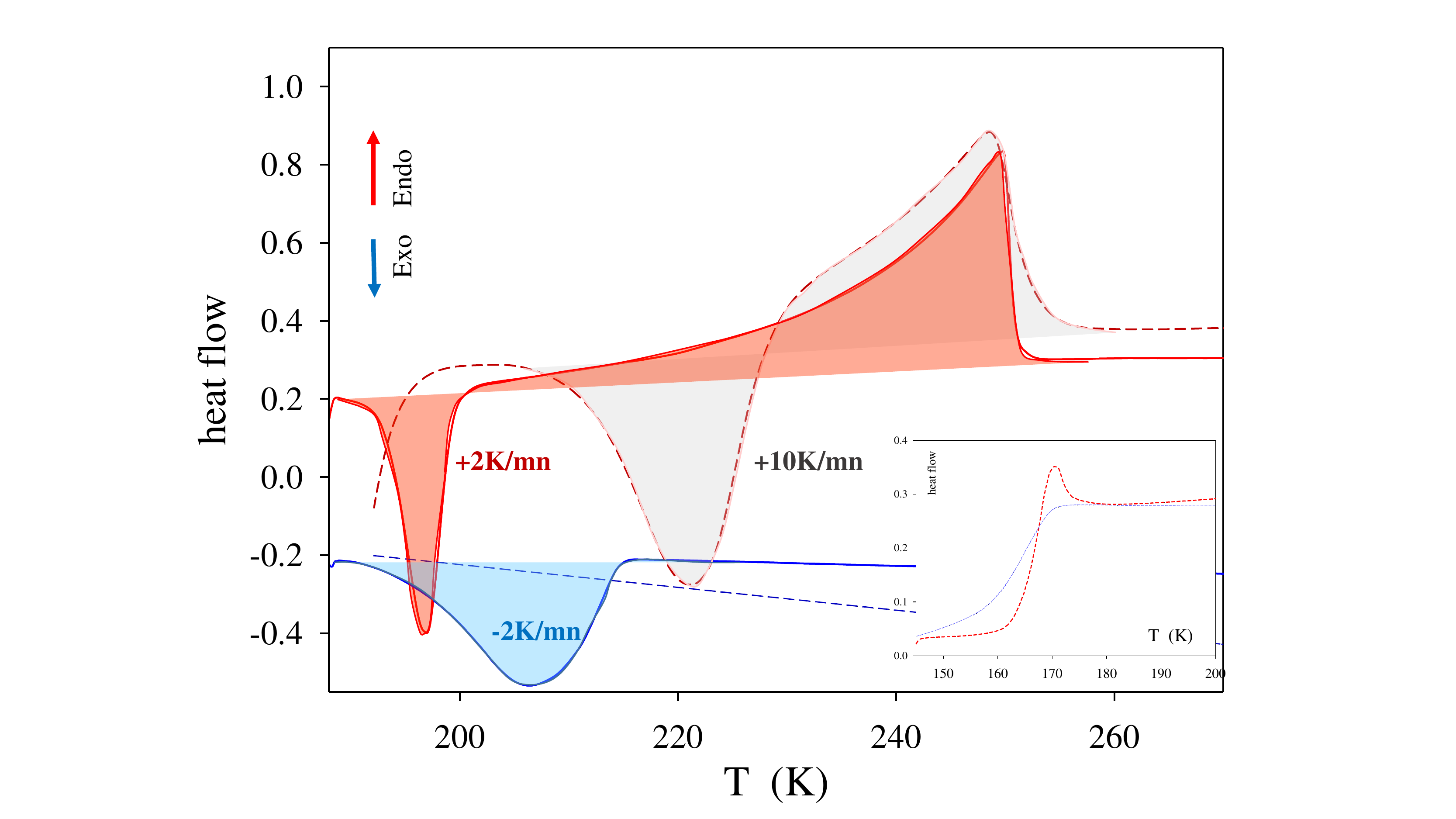}}
\caption{DSC scans of a fully deuterated sample ${\rm C_3D_5(OD)_3+D_2O}$ for $c_g=0.179$ upon cooling (in blue) and heating (in red). The dashed 
lines are for a rate of $10$ K/min and the full lines of $2$ K/min. The areas under the crystallization and the melting peaks correspond to the 
latent heat exchanged during the process. The scans have been rescaled in the figure. Inset: glass transition of the fully deuterated sample with 
cooling/heating rates of $10$ K/min.}
\label{Fig_DSC}
 \end{figure}

Two distinct cooling/heating rates have been applied on a fully deuterated sample with glycerol molar fraction of $c_g=0.179$: see the representative scans in 
Fig.~\ref{Fig_DSC}. At $10$ K/min crystallization is only observed on heating and is immediately followed by melting. From the heat exchanged, one can estimate 
that $26.4\%$ of water crystallizes, while the remaining water is kept trapped in the glycerol matrix (or at the interface with ice crystallites) 
with an estimated glycerol molar fraction of then $22.7\%$. 
After melting, the melted water is redissolved in the mixture. At $2$ K/min, we observe crystallization already on cooling. It is also found on heating, at 
much lower temperatures than for the faster rate. On cooling, $20\%$ of water crystallizes which leads to a remaining solution with a mole fraction 
$c_g\approx 21.3\%$ (and consequently to a glass transition at a molar fraction distinct from the original composition). On heating, an additional $12.7\%$ of 
water crystallizes, an amount that could be considered as representing ``free or interfacial water'' after the glass transition. The remaining glycerol/water liquid mixture 
contains $\sim 23.1\%$ of glycerol. Note that the spread of melting is much larger, almost over $50$ K, for the slower protocol and the maximum of melting 
occurs at a slightly larger temperature. Both aspects illustrate the larger crystallites and wider distribution of crystallite sizes when the kinetics is slow 
(higher $T_m$). The above results are fully compatible with previous ones obtained in~[\onlinecite{hayashi05,popov15}].

The procedures are different than those applied during the structural measurements (see above and main text) and crystallization takes place in 
different temperature ranges. However, the trends shown by the two sets of measurements fully agree. We always observe water (cold) crystallization 
upon heating but crystallization on cooling depend on the cooling rate: no water crystallization for fast enough cooling ($10$ K/min in DSC and down 
to $3$-$6$ K/min in structural measurements) but crystallization for a slower rate of $2$ K/min (DSC).
\\

\section{Estimate of the ice fraction from the high-$Q$ scattered intensity.}

\begin{figure}[tbp]
\centering
\centerline{\includegraphics[width=1.15\linewidth]{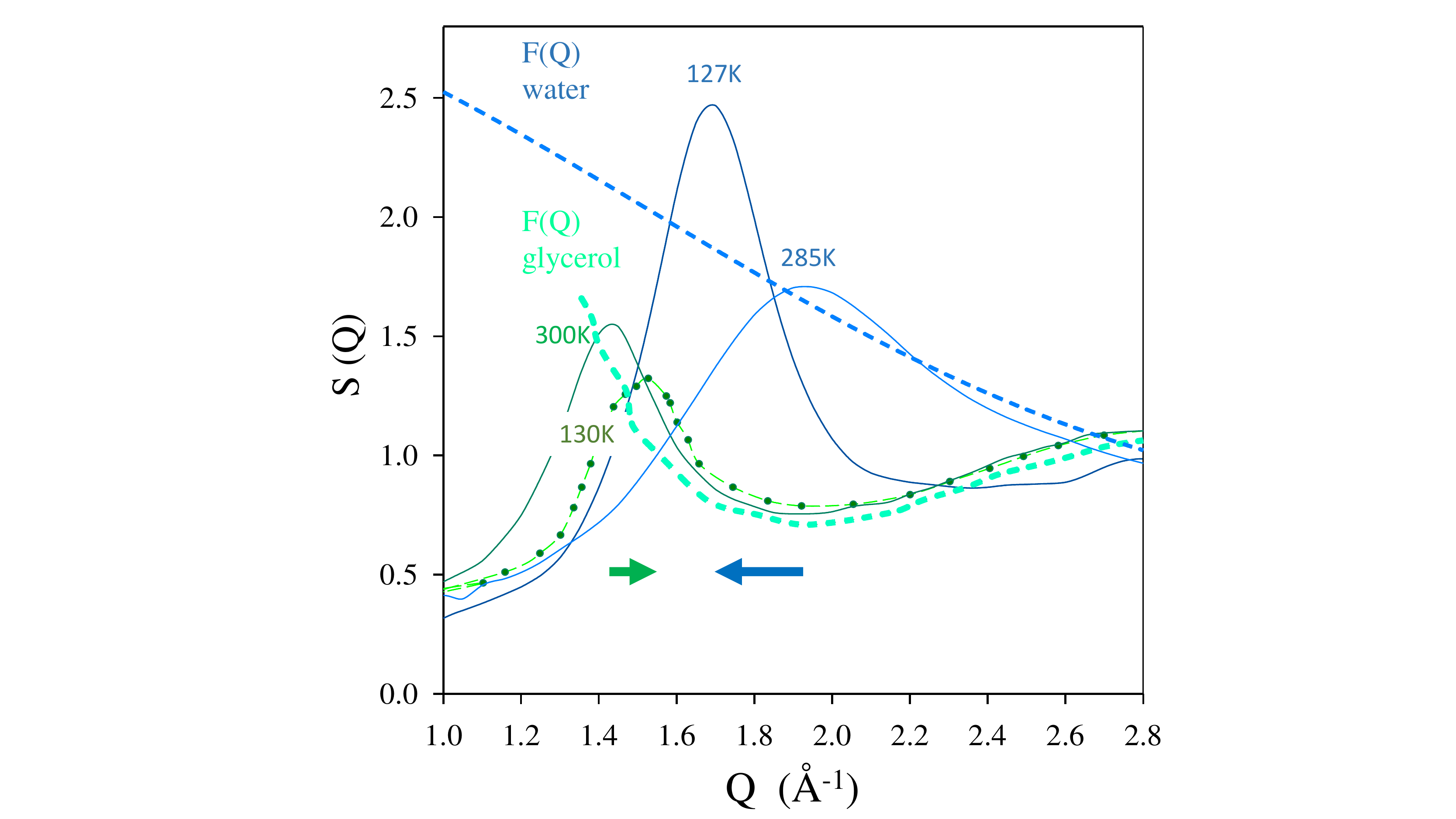}}
\caption{Static structure factor of bulk water at $285$ K (full dark blue line) and at $127$ K in the LDA form (full light blue line)\cite{koza,bellissent}  
and of bulk glycerol\cite{glycerol_denis} at $300$ K (full dark green line) and at $130$ K (full light green line and symbols). The shift of the peak of the $S(Q)$ 
of water to lower $Q$’s (blue arrow) is related to the density decrease of water as $T$ decreases. In contrast, glycerol behaves as a standard molecular liquid 
with a shift to higher $Q$’s (green arrow) related to a density increase.  The dashed lines, respectively green for water and blue for glycerol, are 
the calculated form factor $F(Q)$ in bulk conditions.}
\label{Fig_water-glycerol}
\end{figure}

For a molecular liquid, the static structure factor $S(Q)$ is obtained from the relative positions of the atoms in the sample as the sum of an intramolecular form 
factor, $F(Q)$, and an intermolecular contribution, $D_m(Q)$. $D_m(Q)$ is the sum of all the interatomic components for atoms belonging to distinct molecules 
and its Fourier transform gives the intermolecular pair correlation function. The form factor $F(Q)$ can be calculated from 
the geometry of the molecules: it is the sum of all the pair correlations between atoms of the same molecule. At large $Q$, {\it i.e.}, small distances, only 
the signature of the molecular form factor prevails in $S(Q)$: this is illustrated in Fig. \ref{Fig_water-glycerol} for pure water\cite{koza,bellissent} and 
pure glycerol\cite{glycerol_denis}, where one can see that above $2.4 \AA^{-1}$ the form factor dominates $S(Q)$ for both liquids.  Similarly, for the mixtures 
of the present study, the value of the measured coherent differential scattering cross section around $2.4 \AA^{-1}$ only depends on information at 
the molecular scale. The experimental data around $2.4 \AA^{-1}$ (see Fig. 2 of the main text) corresponds to the coherent differential 
scattering cross section in absolute units 
(barn per steradian and per molecule), which we verify to be equal, before water crystallization takes place, to the calculated value listed in Table~\ref{table1} 
within an uncertainty of $0.5\%$. For each of the two annealing temperatures shown in the upper panels of Fig. 2 (main text), the two red points at $2.4 \AA^{-1}$ 
correspond to the signal of the mixture initially prepared at $c_g=0.178$ before water crystallization (highest point) and to the mixture after partial crystallization 
of water (lowest point). As a result of ice formation, the latter signal contains less water molecules and accordingly a higher glycerol molar fraction $c'_g$. 
The pseudo-molecule used to compute the coherent differential scattering cross section from the data in Table \ref{table1} is thus made of of  [$(1-c_g)$ molecule of 
water + $c_g$ molecule of glycerol] for the initial signal and of [$(1-c'_g)$ molecule of water + $c'_g$ molecule of glycerol] for the final one. The difference between 
the two measured data points at $Q=2.4 \AA^{-1}$ then gives access to the amount of water that crystallizes and to the average glycerol molar fraction of 
the remaining uncrystallized mixture.

At $160$ K, before any annealing, the initial value $\sigma_{{\rm coh,init}}/(4\pi)$ at $2.4 \AA^{-1}$ is $2.045$ barn/(steradian.molecule) with $c_g=0.178$.
At the end of the crystallization process, the final value $\sigma_{{\rm coh,final}}/(4\pi)$ is $1.832$ barn/(steradian.molecule). This lower value is due to the 
decrease of the amount of water molecules in the remaining mixture. The proportion of water in the latter is equal to  
$X_{w,{\rm final}} =(\sigma_{{\rm coh,final}} - c_g\, \sigma_{{\rm coh},g})/\sigma_{{\rm coh},w}$, while the initial proportion is simply 
$X_{w,{\rm init}} =(1-c_g)$ (where $w$ and $g$ refer to water and glycerol, respectively). The fraction of water that crystallizes 
is therefore $(X_{w,{\rm init}}-X_{w,{\rm final}})/ X_{w,{\rm init}}$.  At $160$ K, $21.3\%$ of water crystallizes, and the 
remaining mixture has a glycerol concentration $c'_g=0.216$. There are several sources of uncertainties in this estimation. One may come from the 
uncertainty on the effective deuteration rate of the sample (a $1\%$ difference in the deuteration of glycerol from $99\%$ to $98\%$ leads to an 
error of $0.5\%$ in the estimated fraction of crystallized water). Additional uncertainties can be estimated from the small difference between the 
$Q=2.4 \AA^{-1}$ value in full structure factor at the end of the annealing and that in the amorphous contribution obtained 
after removing the crystalline component determined through the Rietveld analysis (green curve in Fig. 2 of the main text): this difference 
however represents less than $1\%$, which means a $1\%$ uncertainty in the fraction of crystallized water ($22.5\%$ instead of $21.3$). 

With the second thermal treatment corresponding to the slow cooling protocol, crystallization appears rapidly when the sample is 
heated from the glass to $160$ K, and we find that a fraction of $7.5\%$ of water crystallizes.  For the partly deuterated sample 
${\rm C_3H_5(OD)_3+D_2O}$ shown in Fig. 1 of the main text, the proportion of water that has crystallized is $19.6\%$ with a glycerol mole 
fraction of $c'_g=0.212$ in the remaining mixture.

At $170$ K, a similar analysis provides $\sigma_{{\rm coh,init}}/(4\pi)\approx2.043$ barn/(steradian.molecule) with $c_g=0.178$ and 
$\sigma_{{\rm coh,final}}/(4\pi)\approx 1.641$ barn/(steradian.molecule), from which we extract that $39 \pm1\%$ of of water has crystallized with a glycerol 
mole fraction of $c'_g=0.26$ in the remaining mixture. Finally, at $180$ K, $53\%$ of water crystallizes and the glycerol mole fraction in the 
remaining mixture is  $c'_g=0.32$, not far from the concentration estimated from dielectric measurements by Popov {\it et al.}\cite{popov15} for liquid II, 
{\it i.e.}, $0.38$-$0.40$.

The conclusions drawn here disagree with the results of Murata and Tanaka\cite{murata12} who found, when temperature is rescaled by $T_g$, a  
smaller fraction of crystallized water for the fully hydrogenated sample from WAXS experiments. They estimate that the fraction of water that has crystallized is 
about $12\%$ at $162$ K ($T/T_g\approx 1.05$), $24\%$ at $167$ K ($T/T_g\approx 1.08$) and $34$-$44\%$ at $170$ K ($T/T_g\approx 1.10$). This is to 
be compared with what we find here: $21\%$ for $T/T_g \approx 0.97$, $39\%$ for $T/T_g \approx 1.03$, and $53\%$ for $T/T_g \approx 1.09$. These lower estimates 
are possibly due to a less precise estimation procedure, an insufficient annealing time at the lowest temperatures, or an effectively slower cooling rate.
\\

\section{Neutron spin echo (NSE) results for the dynamics.}

\begin{figure}[tbp]
\centering
\centerline{\includegraphics[width=1.05\linewidth]{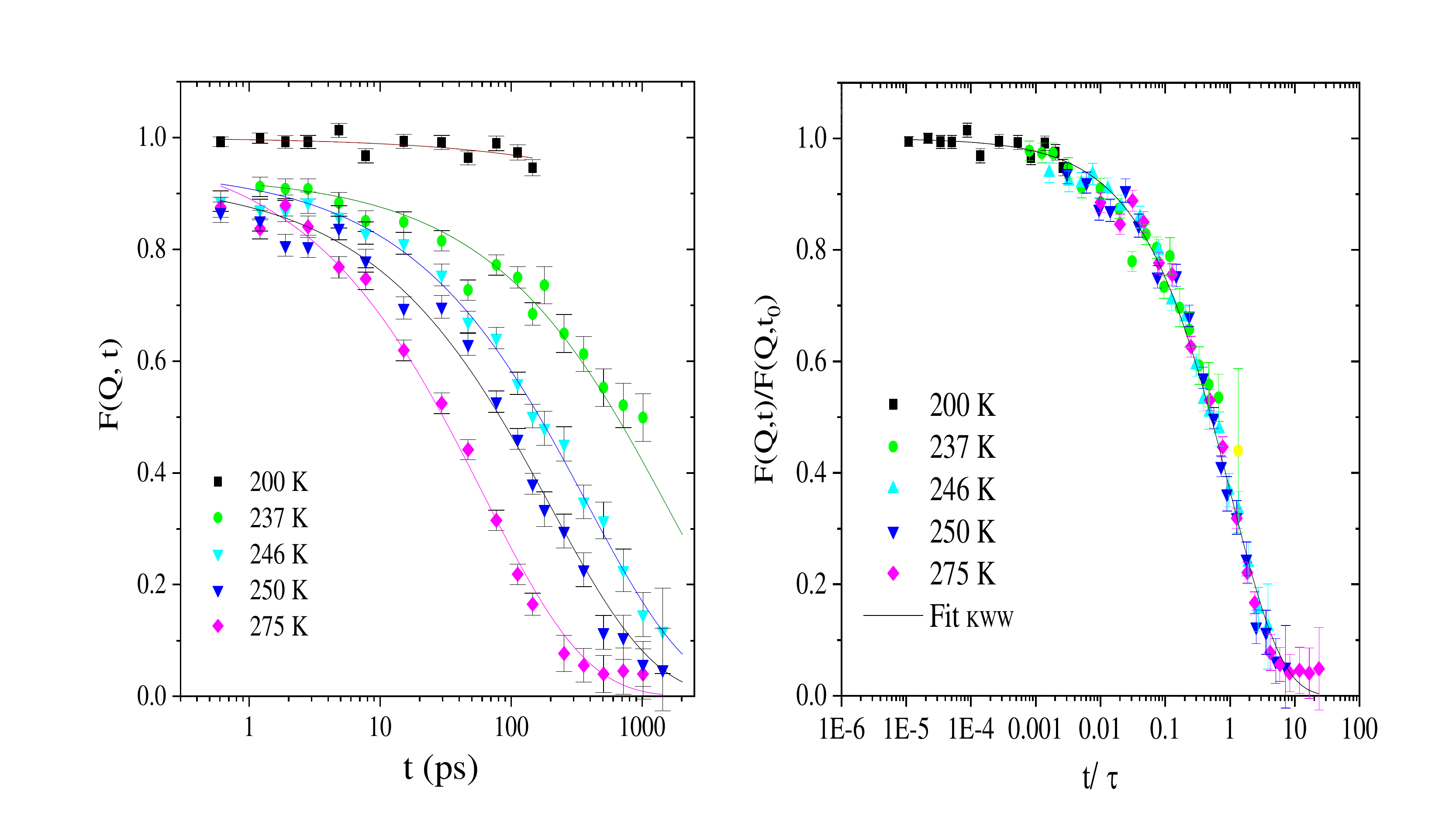}}
\caption{Left: Normalized time-dependent coherent scattering function $F(Q,t)$ obtained by NSE at the wavevector $Q$ of the maximum peak of the static 
structure factor, both in the stable and in the weakly supercooled liquid (which we refer to as liquid I): glycerol-water solution with $c_g= 0.178$. 
Right: Rescaling of the data by using time-temperature superposition; the full line is the best fit to a stretched exponential (KWW function).}
\label{Fig_NSE1}
\end{figure}

The typical neutron spin echo experiment consists in measuring the polarization at (so-called Fourier) times from $t=0$ ps to $2000$ ps. These values are 
then normalized by the signal at $t=0$ and by the resolution function. Typical  curves are shown in Fig.~\ref{Fig_NSE1} (left) for 
$c_g= 0.178$ and for a range of temperatures in the stable and weakly supercooled liquid ($200$ to $280$ K), which we refer to as liquid I, at 
the wavevector $Q=1.9\AA^{-1}$ corresponding to the maximum of the structure factor in this range of temperature. A time-temperature superposition 
curve (right panel of Fig.~\ref{Fig_NSE1}) 
can then be built and fitted with a stretched exponential function (also known as Kohlrausch-Williams-Watts function) to obtain a more robust value of 
the stretching parameter $\beta_{{\rm KWW}}$:  $f(t) = A \exp[-(t/\tau)^{\beta_{{\rm KWW}}}]$. The characteristic time $\tau$ is determined for each temperature 
to provide the best collapse on the master-curve. The stretching exponent is found to be  $\beta_{{\rm KWW}}=0.53$, a value comparable to that of other 
molecular glass-forming liquids in their supercooled regime. The experiments were performed on cooling and are reproducible in the temperature range studied.

We have similarly obtained the NSE data for the solutions with $c_g=0.28$ and $c_g=0.40$.
\\

\section{High resolution $^1$H NMR.}

\begin{figure}[tbp]
\centering
\centerline{\includegraphics[width=1.15\linewidth]{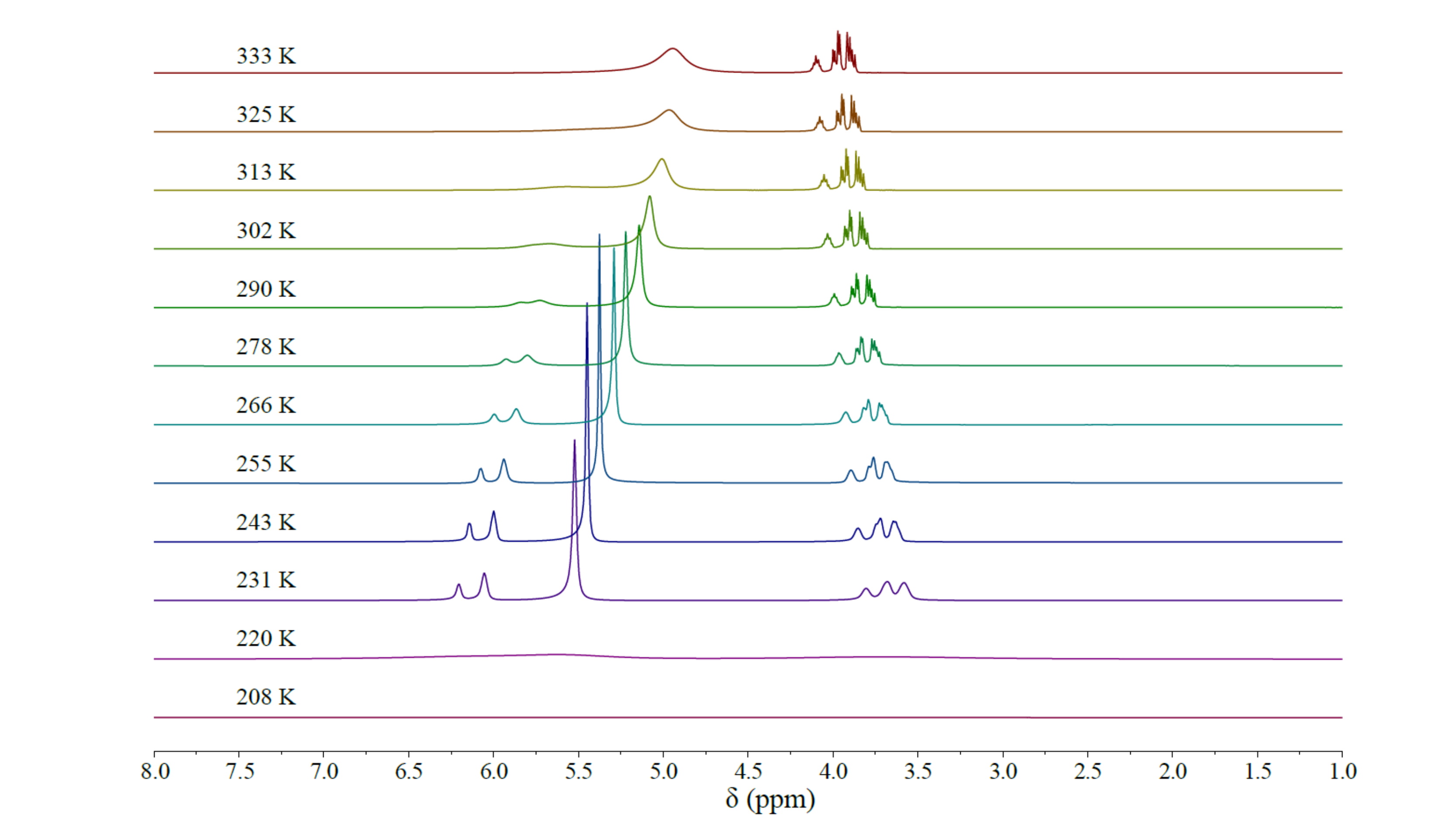}}
\caption{Evolution of $^1$H NMR spectra for the the glycerol/$\rm{H_2O}$ mixture ($c_g=0.191$) as a function of temperature during the cooling stage. 
} 
\label{Fig_NMR1}
\end{figure}

\begin{figure}[tbp]
\centering
\centerline{\includegraphics[width=1.15\linewidth]{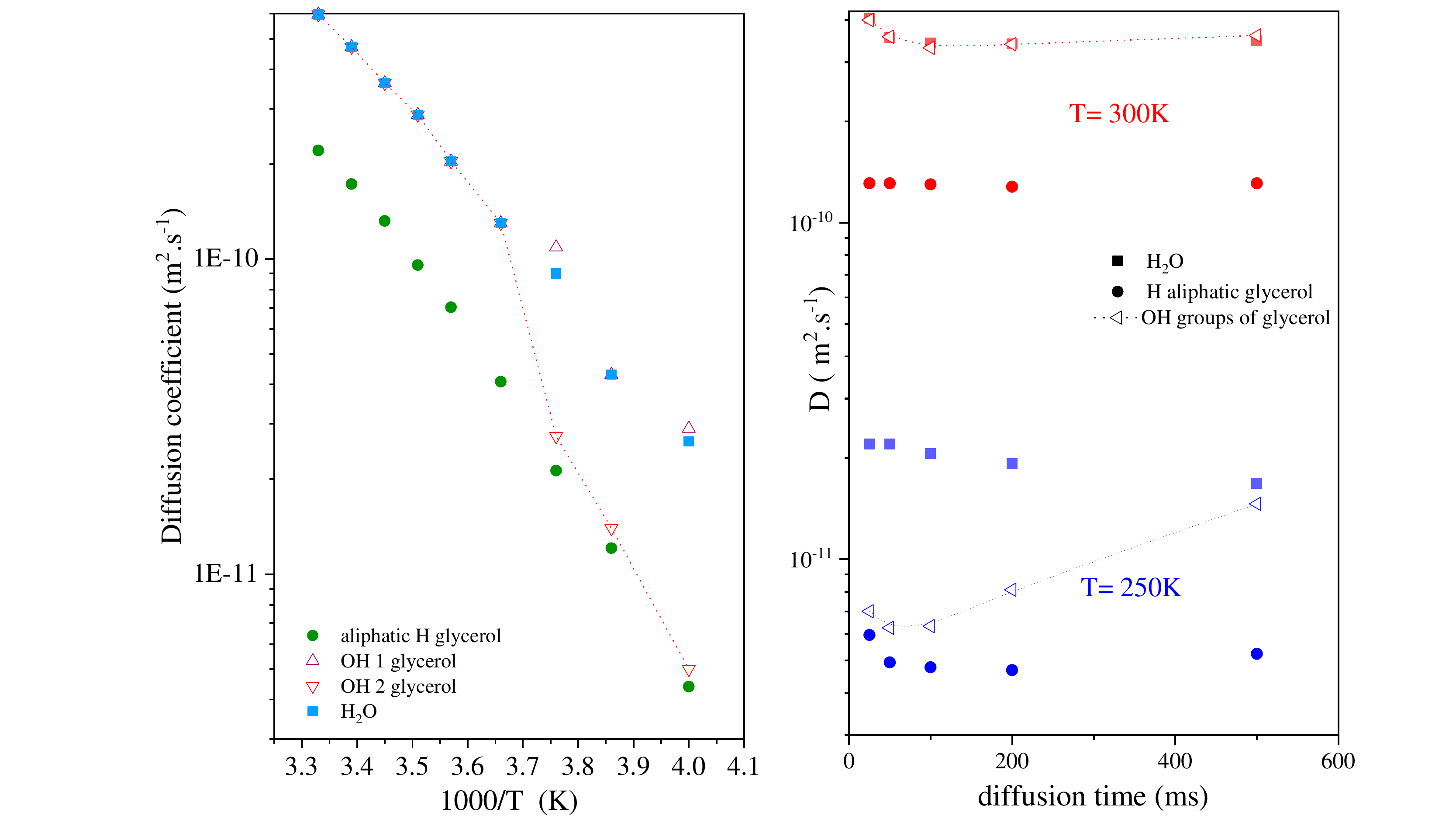}}
\caption{(a) Evolution of the self-diffusion coefficients of the different hydrogen populations (aliphatic for the glycerol backbone, $\rm{H_2O}$ 
and the two populations of glycerol OH) for the glycerol/$\rm{H_2O}$ mixture ($c_g=0.191$) as a function of temperature (measurements performed 
with a diffusion time of $100$ ms). The ratio of the diffusion coefficient of water over that of glycerol increases from $2.7$ at $295$ K  
to $6$ at $250$ K. 
(b) Evolution of the apparent diffusion coefficients with the diffusion time, as measured at $250$ K and $300$ K.}
\label{Fig_NMR2}
\end{figure}

High resolution $^1$H NMR allows one to give a clear complementary picture of the specific behavior of water and glycerol on a temperature 
range that covers the stable and the weakly supercooled liquid regime (liquid I). Fig.~\ref{Fig_NMR1} presents the spectra of the 
glycerol/$\rm{H_2O}$ mixture ($c_g=0.191$) as a function of temperature. At higher temperature, two set of peaks are observed, corresponding to the glycerol 
backbone (around $3.8$ ppm) and to the mobile protons (around $5$ ppm). This unique peak for the OH groups of glycerol and $\rm{H_2O}$ molecules is the 
signature of a rapid exchange regime between all these exchangeable hydrogen atoms. However, when the sample temperature is decrease below 
$313$ K, this signal splits into two components, corresponding respectively to the OH groups of glycerol (above $6$ ppm) and water 
(around $5.5$ ppm). This results from a slowdown of the exchange regime between the groups belonging to glycerol and water. At even lower 
temperature ($280$ K), the two types of OH group of glycerol are also clearly split, which means an even longer residence time of all types of OH's 
on the glycerol molecules. Going to even lower temperatures, the signal broadens, indicating a strong decrease of molecular mobility. This 
phenomenon is reversible when heating the sample, with however a temperature hysteresis of around $20$ K. In this domain, peak integration 
shows that half of the signal of water is lost compared to higher temperature, while the signal 
of glycerol is not affected. This is due to the selective crystallization of a fraction of the water in this mixture below $240$ K. Further information can be gained 
from the measurement of the self-diffusion coefficients by pulsed-field gradient NMR. The specific behavior of each hydrogen group can be selectively measured. 
The variation of the self-diffusion coefficients with temperature is presented in Fig.~\ref{Fig_NMR2}: one can see that the ratio of the diffusion 
coefficient of water over that of glycerol increases from $2.7$ at $295$ K  to $6$ at $250$ K.  At high temperature, the self-diffusion coefficients of all the 
OH groups have similar values, significantly higher than that of the glycerol backbone. Note finally that the ratio between water and glycerol diffusion coefficients 
at $295$ K is comparable to the value of $2.5$ found for a fully deuterated sample at $298$ K\cite{NMR_deuterated}.

Around $265$ K, one can notice that a fraction of the OH groups has the same diffusion coefficient as the glycerol backbone, indicating their 
longer residence time on glycerol. This slowing-down of the exchange rate is also seen in Fig.~\ref{Fig_NMR2} where the apparent diffusion coefficients 
are measured as a function of the diffusion time at two different temperatures. At high temperature, no variations are measured for diffusion time between 
$20$ ms and $500$ ms, indicating a Fickian-like behavior. However, at $250$ K, the apparent diffusion coefficient of hydroxyl groups is strongly dependent 
on the diffusion time. At shorter time it is similar to that of the glycerol backbone, while at longer time it is water-like. This characterizes an exchange regime 
in the range of hundred of ms. The timescale over which nano-segregation locally persists therefore strongly increases with decreasing temperature.

\section{Structure of liquids I and I'.}

\begin{figure}[tbp]
\centering
\centerline{\includegraphics[width=1.1\linewidth]{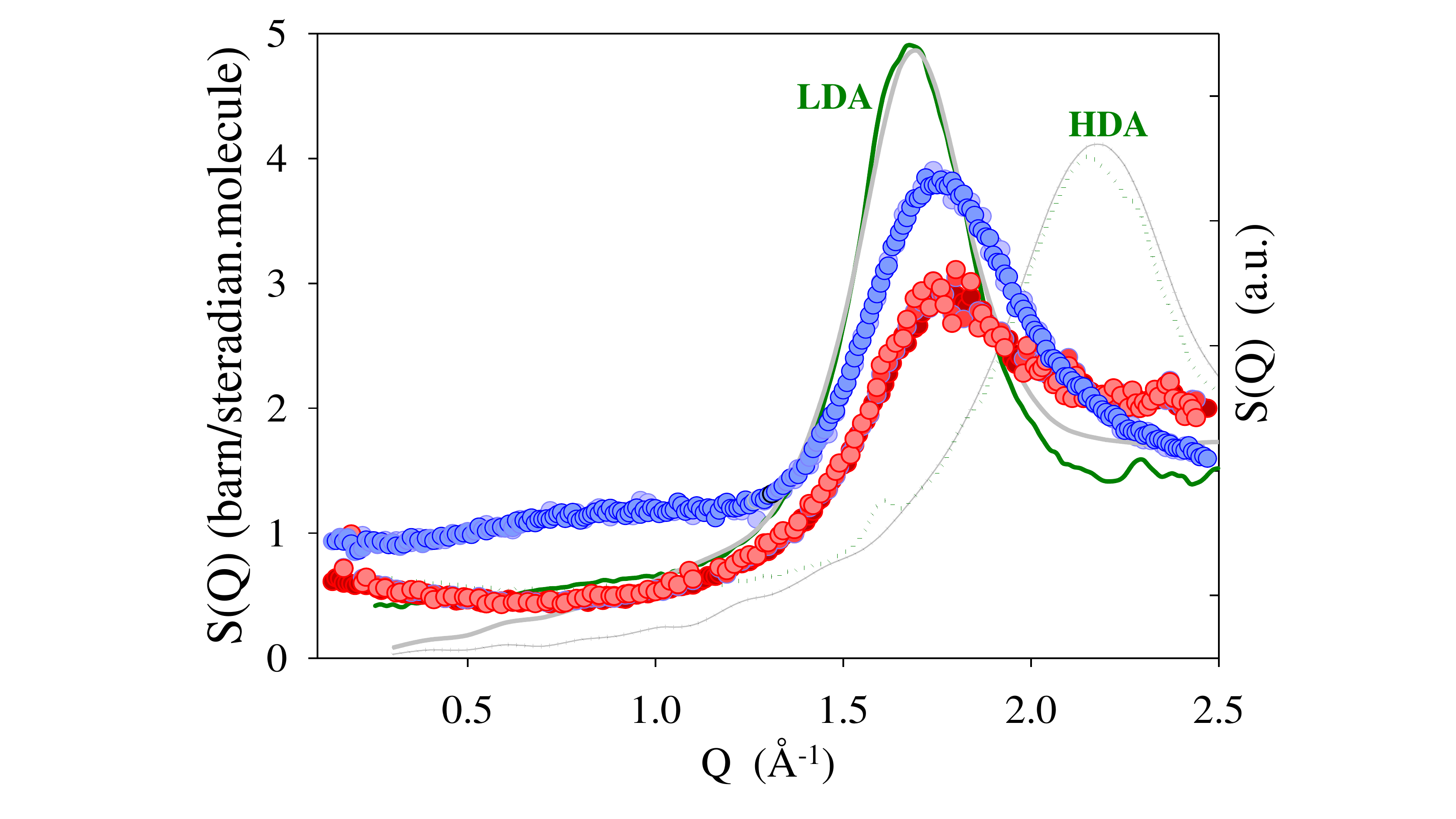}}
\caption{Comparison between the static structure factor $S(Q)$ of liquid I' obtained by neutron scattering at $130$, $160$ and $170$ K (blue symbols: 
${\rm C_3H_5(OD)_3+D_2O}$; red symbols: ${\rm C_3D_5(OD)_3+D_2O}$) and that of the HDA at $77$ K and the LDA at $120$ K and $127$ K (right axis, 
in arbitrary units; data from~[\onlinecite{koza,bellissent}]).}
\label{Fig_comparisonHDA-LDA}
\end{figure}

First we show that water in the nano-segregated liquid I' bears no resemblance to its high-density amorphous form and rather look like the low-density form: 
This is unambiguously seen from Fig. \ref{Fig_comparisonHDA-LDA}.

We also provide additional neutron-scattering structural data illustrating the difference between what we refer to as liquid I and liquid I' for the glycerol-water 
solution with $c_g\approx 0.18$.

\begin{figure}[tbp]
\centering
\centerline{\includegraphics[width=1.15\linewidth]{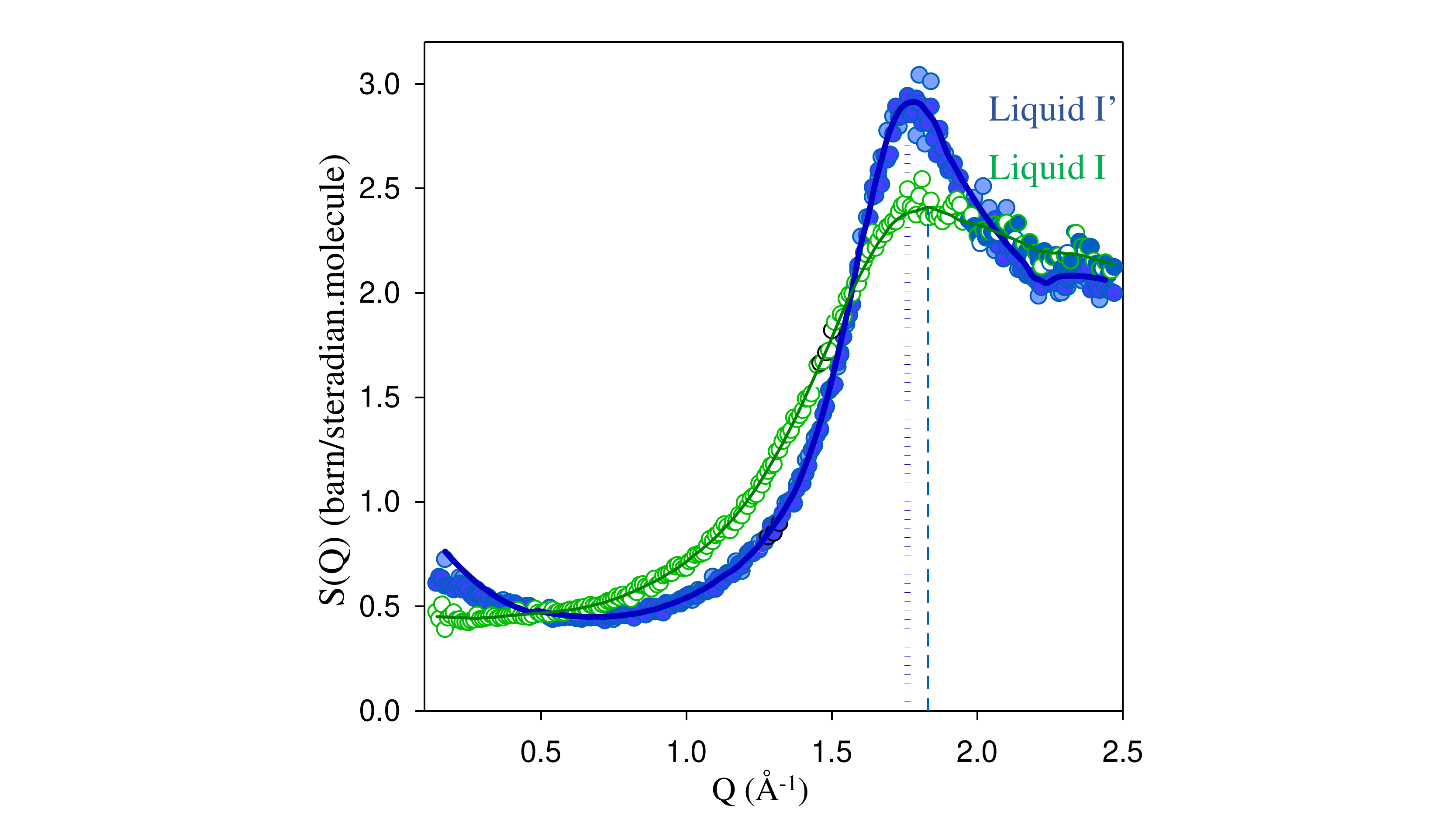}}
\caption{Structure factor $S(Q)$ of liquid I at $260$ K above melting (green symbols) and of glass/liquid I' at $130$ K and $160$ K 
(blue symbols) for the $c_g=0.178$ glycerol-water solution. This illustrates the increased nano-segregation effect during cooling.}
\label{Fig_nanosegreg}
\end{figure}


\begin{figure}[tbp]
\centering
\centerline{\includegraphics[width=1.3\linewidth]{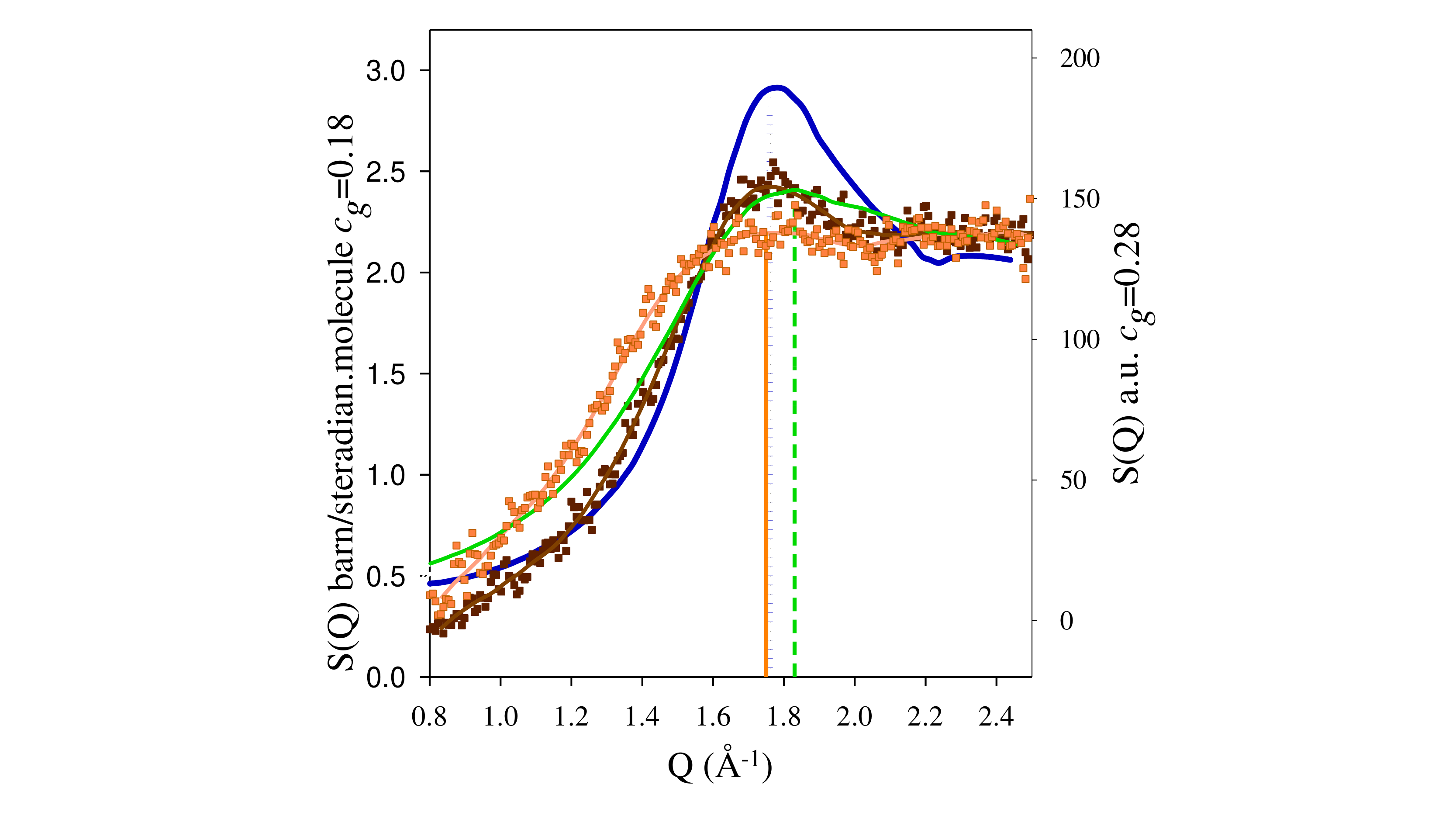}}
\caption{Comparison of the structure factors of liquid I (green line) and liquid I' (blue line) for $c_g=0.178$ (same data as in Fig.~\ref{Fig_nanosegreg})
with those of liquid I (orange squares and associated line) at $275$ K and of glass/liquid I' at $130$ K (brown squares and associated line) 
for a higher glycerol concentration $c_g= 0.28$.}
\label{Fig_two_cg}
\end{figure}

In Fig.~\ref{Fig_nanosegreg} we display the structure factor $S(Q)$ of liquid I at $260$ K above melting  and of that of liquid/glass I' at $130$-$160$ K 
(prior to water crystallization). The maximum shifts from $1.83 \AA^{-1}$ at $260$ K to $1.75 \AA^{-1}$ at $130$ K, a trend which is typical of bulk water 
(see Fig.~\ref{Fig_water-glycerol}), and one can see an intensity increase at lower $Q$'s  at $170$ K. These features are indications of an 
increased nano-segregation of water during the fast cooling (and during the subsequent heating to the annealing temperature, prior to ice 
formation). Further support comes for the fact that while the main peak follows the trend of bulk water the average liquid density conforms to the 
behavior of a conventional molecular liquid and increases between the high-temperature liquid I and the low-temperature liquid I'\cite{density}.

Finally, we also show a comparison of the evolution of the structure between liquid I (above melting) and glass/liquid I' at $c_g\approx 0.18$ and at a higher 
glycerol concentration $c_g= 0.28$. Although a detailed comparison is difficult because of the change in the weighting of the partial structure factors between 
$c_g\approx 0.18$ and $c_g=0.28$, one can notice that the position of the main peak of liquid I' is around $1.75\AA^{-1}$ and does not seem to change much 
with concentration. The variation with temperature ({\it i.e.}, between liquid I and liquid I') of the peak position is significant and goes to lower $Q$ as $T$ decreases 
for $c_g\approx 0.18$ whereas it is negligible for $c_g=0.28$. The displacement to lower $Q$ tracks the nano-segregation of water, and this effect disappears as $c_g$ 
increases. We conjecture that at even higher glycerol concentration, for $c_g\gtrsim 0.38$, the peaks shifts to higher $Q$ as $T$ 
decreases, as seen in standard molecular liquids (see glycerol in Fig.~\ref{Fig_water-glycerol}).
\\


\begin{thebibliography}{ÇÇ}
\bibitem{footnote_avrami}
For the same value of the exponent and still assuming that the growth is diffusion-limited it could also be that the grains are spherical but that the nucleation rate 
changes with time. This is not crucial to our analysis.
\end{thebibliography}
\end{document}